\pdfoutput=1
\documentclass[12pt]{article}
\usepackage[margin=1in]{geometry}
\usepackage{amsmath}
\usepackage{titling}
\usepackage{blindtext}
\usepackage{pgfplots}
\usepackage{natbib}
\usepackage{comment}
\usepackage{setspace}
\usepackage{authblk}
\usepackage[colorlinks=true,linkcolor=blue,allcolors=blue]{hyperref}   
\usepackage{url}  
\usepackage[percent]{overpic}
\usepackage[section]{placeins}
\usepackage{pbox}
\usepackage{comment}
\usepackage{titling}
\usepackage{natbib}
\usepackage{soul}
\usepackage{siunitx}
\usepackage{textcomp}
\newcommand{\GG}[1]{}

\usepackage{ragged2e}
\justifying

\usepackage{comment}
\usepackage[percent]{overpic}
\usepackage[flushleft]{threeparttable}
\usepackage[clockwise, figuresright]{rotating}
\usepackage{pbox}
\usepackage{multirow}
\usepackage{longtable}
\usepackage{enumitem}

\def\soho{{\sl SOHO}}
\def\stereo{{\sl STEREO}}
\def\sdo{{\sl SDO}}

\begin{document}

\title{Ensemble simulations of the 12 July 2012 Coronal Mass Ejection with the Constant Turn Flux Rope Model}

\author[1]{Talwinder Singh}
\affil[1]{Center for Space Plasma and Aeronomic Research, The University of Alabama in Huntsville, AL 35805, USA}

\author[1]{Tae K. Kim}

\author[1,2]{Nikolai V. Pogorelov}
\affil[2]{Department of Space Science, The University of Alabama in Huntsville, AL 35805, USA}

\author[3]{Charles N. Arge}
\affil[3]{Solar Physics Lab, NASA/GSFC, Greenbelt, MD 20771, USA}

\setcounter{Maxaffil}{0}
\renewcommand\Affilfont{\itshape\small}
\date{}  

\begin{titlingpage}
    \maketitle
    
\begin{abstract}
Flux-rope-based magnetohydrodynamic modeling of coronal mass ejections (CMEs) is a promising tool for prediction of the CME arrival time and magnetic field at Earth. In this work, we introduce a constant-turn flux rope model and use it to simulate the 12-July-2012 {16:48} CME in the inner heliosphere. We constrain the initial parameters of this CME using the graduated cylindrical shell (GCS) model and the reconnected flux in post-eruption arcades. We correctly reproduce all the magnetic field components of the CME at Earth, with an arrival time error of approximately 1 hour. We further estimate the average subjective uncertainties in the GCS fittings, by comparing the GCS parameters of 56 CMEs reported in multiple studies and catalogs. We determined that the GCS estimates of the CME latitude, longitude, tilt, and speed have average uncertainties of $5.74^\circ$, $11.23^\circ$, $24.71^\circ$, and 11.4\% respectively. Using these, 
we have created 77 ensemble members for the 12-July-2012 CME. We found that 55\% of our ensemble members correctly reproduce the sign of the magnetic field components at Earth. We also determined that the uncertainties in GCS fitting can widen the CME arrival time prediction window to about 12 hours for the 12-July-2012 CME. On investigating the forecast {accuracy} introduced by the uncertainties in individual GCS parameters, we conclude that the half-angle and aspect ratio have little impact on the predicted magnetic field of the 12-July-2012 CME, whereas the {uncertainties} in longitude and tilt can introduce relatively large {spread} in the magnetic field predicted at Earth.
\end{abstract}
\end{titlingpage}

\section{Introduction}

Coronal mass ejections (CMEs) are plasma and magnetic field structures that usually erupt from highly magnetic, closed-field regions of the Sun called active regions (ARs). {CMEs can carry large kinetic and magnetic energy with their erupted mass ranging between $10^{9}$ kg and $10^{13}$ kg \citep{Vourlidas10}, and speeds that may exceed 3000 km/s.} An Earth-directed CME carrying a negative out-of-the-ecliptic  magnetic field component ($B_\mathrm{z}$) can cause extreme space weather events that can affect space-borne and ground-based technological systems. Therefore, predicting CME arrival times and magnetic field values at Earth by properly treating their interplanetary propagation is of immense importance.

A number of CME propagation models have been developed over the years to reproduce their travel through interplanetary space and forecast arrival times at Earth. Such models include empirical models \citep[e.g.][]{Vandas1996, Brueckner1998, Gopalswamy2001, Gopalswamy2005, Wang2002}, drag based models \citep{Vrsnak2002}, and physics-based models, such as the Shock Time of Arrival (STOA) \citep{Dryer1984}, STOA-2 \citep{Moon2002}, and HAFv.2 \citep{Fry2001}. Newer-generation prediction models such as Wang-Sheeley-Arge (WSA)-ENLIL-Cone model \citep{Odstrcil04} use magnetohydrodynamic (MHD) simulations in which CMEs are inserted into the ambient flow as non-magnetized over-pressured plasma blobs. {The accuracy of this type of models is not much higher than in the previously-mentioned models of lesser complexity and have a mean absolute error of more than 10 hr \citep{Riley2018, Amerstorfer18}.} \citet{Gopalswamy2013} have shown that the performance of the ENLIL-cone model is similar to that of the empirical shock arrival model \citep{Gopalswamy2005}. {The ENLIL-cone and HAFv.2 models can calculate magnetic field in the ICME sheath region because a preceeding shock distorts the interplanetary magnetic field. However, they are unable to describe magnetic field in ICME clouds at Earth because of the absence of an explicitly defined magnetic field inside them \citep{Fry2001,Odstrcil04}.} Attempts have been made to remedy this by adding a simple magnetic field into the plasma blobs \citep[see, e.g.,][]{Shen11}.

According to the standard model of CME eruption, also called the CSHKP model \citep{Carmichael1964, Sturrock1968, Hirayama1974, Kopp1976}, pre-eruptive states are characterized by the presence of an axial magnetic flux, which may be either a sheared arcade or a flux rope located above the magnetic polarity inversion line separating positive and negative radial magnetic fluxes in an active region. Such flux ropes can rise due to different magnetic force imbalances, and the overlying magnetic field lines add the majority of poloidal flux to the flux rope during an eruption \citep[see the review of][]{Chen11}. The interplanetary counterparts of CME flux ropes which are characterized by a large, rotating magnetic field, as revealed by in-situ measurements, are called the magnetic clouds \citep{Burlaga81}.

Using in-situ observations of ICMEs between 1978–1982, \citet{Gosling90} showed that $\sim30\%$ of ICMEs sampled at 1 AU had magnetic clouds associated with them. However, \citet{Marubashi00} later claimed that up to $80\%$ of all ICMEs involve magnetic clouds. A recent study by \citet{Song20} argues that all CMEs observed at 1 AU have magnetic clouds, though some are not directly observed because the observing spacecraft crosses the flanks of the ICME. Studies, such as \citet{Gopalswamy2018a} and \citet{Sarkar2020} show the key importance of flux ropes in the development of forecast models capable of predicting the CME magnetic properties at 1 AU. Therefore, the new generation of CME forecast models is strongly associated with the advanced description of flux rope structures \citep{Shen11,Shiota16, Isavnin16, Vandas17, Jin17b, Scolini19, Singh20b}. 

MHD simulations of the solar wind (SW) and flux-rope-based CMEs can either start in the lower corona \citep{Manchester04b, Jin16, Singh18} or the inner heliosphere (IH) \citep{Verbeke19, Singh20b}. If a simulation starts in the lower corona, one can allow a CME to be automatically generated in the ambient solar corona background \citep{Amari14}. Otherwise, an analytically-generated flux rope can be inserted \citep{Titov99, Gibson98, Singh20a}. The latter approach is less computationally expensive than the former and allows for better constraining of magnetic flux inside flux-rope models \citep{Singh19}. The approach where CMEs are inserted in the IH is computationally less expensive and easier to implement since the boundary condition treatment in the superfast magnetosonic flow is mathematically trivial and no coronal heating model is required. However, this approach can lead to errors associated with the extrapolation of the SW and CMEs from the lower corona to the inner boundary of the IH models. This approach is more common to the current operational ICME predictions such as by NASA Goddard Space Flight Center/Space Weather Research Center, NOAA/Space Weather Prediction Center USA, Australian Bureau of Meteorology, Korean Space Weather Center, and UK Met Office, which use the WSA-ENLIL-Cone model. This approach is also being pursued by several other models, e.g., EUHFORIA \citep{Verbeke19} and the one described by \citet{Shiota16}.

In an IH model, a CME can either be inserted step by step at the inner boundary \citep{Verbeke19} or in a single step above the inner boundary \citep{Singh20b}. If a CME flux rope is inserted by superimposing an analytic model either in the corona or in the IH, its physical and magnetic properties should be driven by observations for accurate forecasting. These properties include the CME direction, tilt, shape, speed, mass, poloidal and toroidal magnetic fluxes, and helicity sign. {The CME tilt refers here to the angle between the line connecting two footpoints of the CME flux rope and the equatorial plane.} This tilt can be seen as angle $\gamma$ in Figure 1 of \citet{Thernisien09}. Moreover, the ambient SW through which this CME will propagate should also be based on data-driven models. CME observations used to constrain any model are typically accompanied by uncertainties in different measured quantities. It is crucial to understand how these uncertainties propagate through the computational region and affect the model forecasts at Earth. Ensemble modeling of CMEs is a viable approach to the study of the propagation of these {uncertainties} and ensure the probabilistic forecasting \citep{Pizzo15, Mays15, Murray18, Amerstorfer20}.  

In this paper, we use the constant-turn flux rope model to simulate a CME that erupted on 12 July 2012 at {16:48}. The geometry of this model is described by the FRiED model of \citet{Isavnin16}. We use the analytic formulae governing the constant turn flux rope, as described by \citet{Vandas17}, to specify the magnetic field in this model. We estimate the 12 July 2012 CME direction, tilt, half-angle, aspect ratio, and speed using the graduated cylindrical shell (GCS) model and initialize our flux rope using these properties. The magnetic flux associated with a CME in the flux-rope model is calculated using the poloidal flux estimated from post-eruption arcades (PEAs).

We demonstrate that by using these observational initial conditions it is possible to correctly reproduce the sign of all three magnetic field components of the magnetic cloud of this ICME. The arrival time, density, and speed are also reproduced with reasonably good accuracy as compared to previous attempts to simulate this CME using different models \citep{Shen14, Scolini19, Singh20b}. {We also perform an analysis of differences in GCS fittings results reported in multiple publications and online catalogs to quantify the uncertainties associated with such fittings.} We further investigate how the uncertainties in the GCS parameters, both individually and collectively, affect the simulated CME properties at Earth by using an ensemble modeling of the 12 July 2012 CME.

In Section~\ref{sec:Models}, we describe the GCS model, SW model, and the flux-rope-based CME model. We then show our CME simulation and GCS {uncertainty} analysis results in Section~\ref{sec:Results}, followed by the conclusions in Section~\ref{sec:Conclusions}.

\section{Data and Models}\label{sec:Models}
{In this work, we use the level~0.5 image data {collected on 12 July 2012 between 16:24 and 18:54 UT} from the Sun-Earth Connection Coronal and Heliospheric Investigation (SECCHI)/Cor2 ~\citep{Howard08} coronagraphs on board the Solar Terrestrial Relations Observatory (\stereo) A \& B~\citep{Kaiser08} spacecraft and process them to level~1 using the \textit{secchi\_prep} program in the IDL SolarSoft library, which converts the units from data numbers (DN) into Mean Solar Brightness (MSB). We also use level~1 image data {collected on 12 July 2012 between 16:24 and 17:18 UT} from the Large Angle Spectroscopic Coronagraph (LASCO)/C2/C3~\citep{Brueckner95} coronagraph of the Solar and Heliospheric Observatory (\soho) spacecraft. Level~1 data for both these instruments are suitable for quantitative scientific analysis.} The coronagraph data are used to estimate the speed, direction, tilt, half-angle, aspect ratio, and speed of the 12 July 2012 CME. The model used to estimate these values is described in Section~\ref{sec:GCS}. We use the solar extreme ultraviolet (EUV) {images in 94, 131, and 193 \AA\ wavelengths captured on 12 July 2012 at 22:30 UT and line of sight (LOS) magnetic field observations at 12 July 2012 16:10 UT} from  Solar Dynamics Observatory's (\sdo) Atmospheric Imaging Assembly (AIA) \citep{Pesnell12} and Helioseismic and Magnetic Imager (HMI) \citep{Schou12, Hoeksema14}, respectively. {We use the \textit{aia\_prep} program in SolarSoft to read in and calibrate AIA level~1 data to level~1.5. Level~1 data include bad-pixel removal, de-spiking, and flat-fielding. Additional roll-corrections applied to Level~1.5 data make the solar north direction vertical in the images, re-scale  the images to $0''.6$  pixels, and translate them to match  the solar disk and image centers. The HMI data are read using \textit{hmi\_prep} program in SolarSoft, which processes  magnetograms to ensure correct roll angles, and translates them to put the solar disk center at a magnetogram center.} These data are used to estimate the magnetic flux and helicity sign of the 12 July 2012 CME. The methods used to estimate these values are described in Section~\ref{sec:July12CME}. We use  1 hour averaged data provided by NASA/GSFC's OMNI data set through OMNIWeb \citep{King05} to compare the in-situ properties at Earth with our simulations of 12 July 2012 CME.

\subsection{Graduated Cylindrical Shell model}\label{sec:GCS}
\begin{figure}[!htb]
\center
\includegraphics[scale=0.1,angle=0,width=14cm,keepaspectratio]{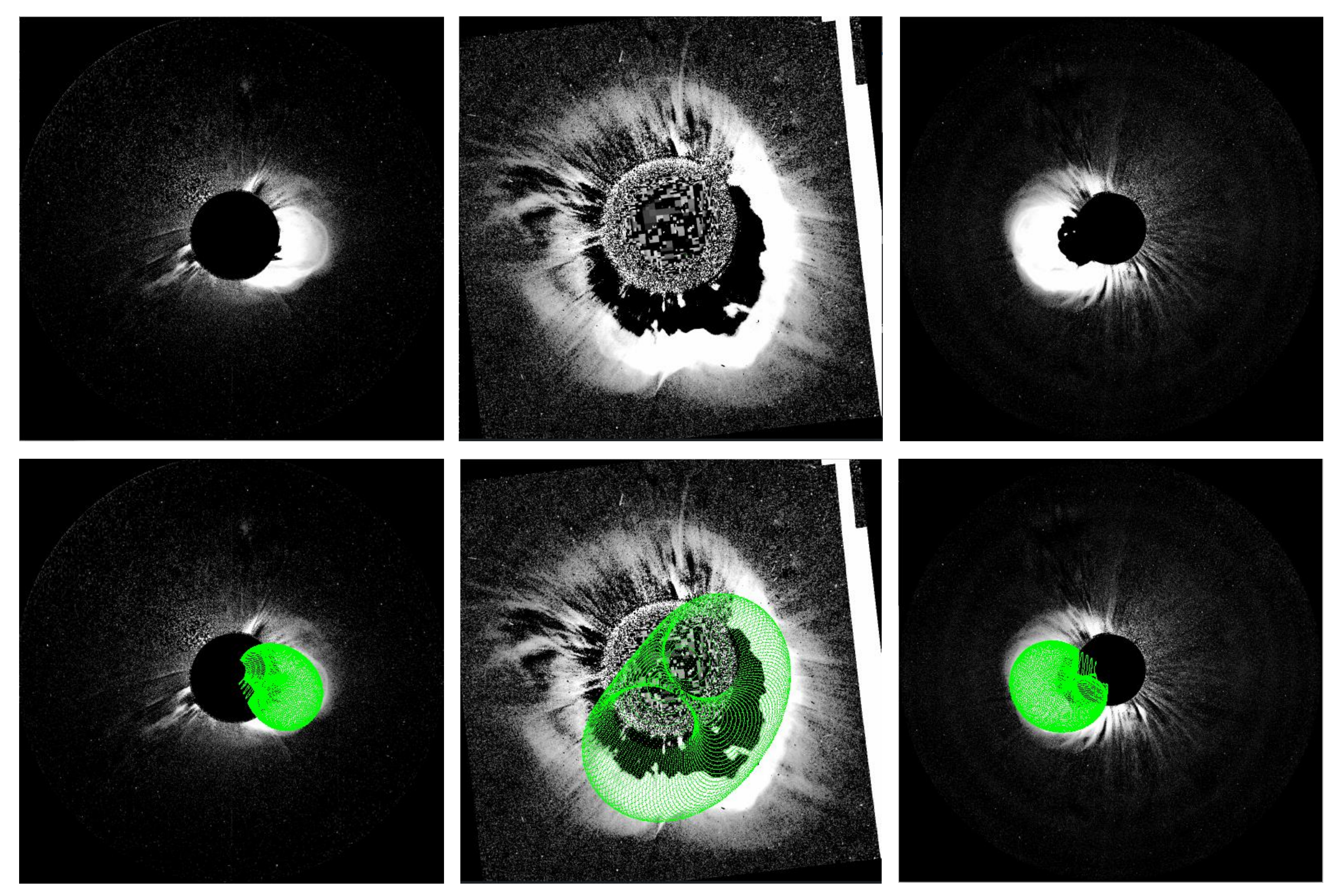}
\caption{(\textit{Top panel, from left to right})The  July 12, 2012 CME seen in STEREO B Cor2, SOHO C2 and STEREO A Cor2 coronagraphs respectively. (\textit{Bottom panel}) The same CME fitted with the GCS model.}
\label{GCS}
\end{figure}

The Graduated Cylindrical Shell (GCS) model uses three different views of a CME from \stereo\,-A \& -B, and \soho \, coronagraphs to visually fit the observed CME with a simplified structure consisting of conical legs and a curved front, {resembling a hollow croissant with a circular cross-section.} The GCS method has been implemented in IDL using the \textit{rtsccguicloud} program~\citep{Thernisien06}. An example of GCS fitting using this tool is shown in Figure~\ref{GCS}, for the July 12, 2012 CME. The GCS model has six free parameters, namely, the latitude, longitude, tilt, height, half-angle, and aspect ratio, which can be modified to fit the 3D flux rope structure in all three images simultaneously. {The \textit{rtsccguicloud} program allows for this modification by using sliders to modify each parameter independently. The mathematical description of the geometry of this model is described in detail by \citet{Thernisien11}.} Notice that the CME height, being one of the GCS parameters, can be converted to speed by fitting a linear function to the height-time graph. {We decided to use linear fitting instead of quadratic fitting because the GCS uncertainty analysis we perform later in the study is based on the catalogs that all reported only the linear speeds.} One cannot reliably derive GCS parameters with a single coronagraph image, since the majority of structural information of a CME is lost when its 3D shape is projected onto a 2D plane (plane of the sky). The availability of multiple images made from different viewpoints is critical for our ability to remove these projection effects. However, CMEs typically have complicated shapes that are difficult to fit with the idealized shapes assumed by the GCS model. Therefore, the CME parameters estimated by the GCS model can have significant uncertainties. We will discuss them in more detail in Section~\ref{sec:GCS_uncertainity}.

\subsection{Magnetohydrodynamic models}
In this study, we perform MHD modeling using the Multi-Scale Fluid-Kinetic Simulation Suite (MS-FLUKSS), a  collection of highly parallel modules capable of performing adaptive mesh refinement (AMR) simulations of the SW in the presence of neutral atoms, nonthermal ions, and turbulence, etc.  \citep{Pogorelov14}. We describe our SW and CME models in the following subsections.

\subsubsection{Solar wind model}
We simulate the ambient SW in the IH by solving the ideal MHD equations with the finite-volume, total variation diminishing (TVD) approximation on a non-uniform spherical grid with 150, 256, and 128 cells in the $r$, $\phi$, and $\theta$ direction, respectively. We set the inner and outer boundaries of the spherical domain at 0.1 and 1.5 AU, respectively. The IH model uses a time-series of WSA maps as inner boundary conditions at 0.1 AU \citep[e.g.,][]{Kim19}. We employ the Air Force Data Assimilative Photospheric Flux Transport (ADAPT) synchronic maps using the NSO/GONG magnetograms \citep{Arge10,Arge11,Arge13,Hickmann15} as input to the WSA model at the solar surface. The WSA model extrapolates the photospheric magnetic field to a spherical source surface at 2.5 $R_\odot$ using the potential field source surface (PFSS) model, and then to the WSA outer boundary at 0.1 AU using the Schatten current sheet model \citep{Schatten71}. The WSA model also calculates the SW speed at its outer boundary as a function of the flux expansion factor and distance to the nearest coronal hole boundary \citep{Arge03, Arge05, McGrogor2011}. To generate the background SW for the flux-rope CME, we use the best ADAPT-WSA realization (out of 12), based on their comparison with near-Earth SW data.

We interpolate the WSA maps, both in time and space, from the original $2.5^\circ \times 2.5^\circ$ resolution to the IH model base grid of $\sim 1.4^\circ \times 1.4^\circ$. We also scale the WSA magnetic field strengths by a factor of 3 during this process to compensate for the systematic underestimation of the open magnetic flux at 1 AU \citep{Linker16, Linker17, Wallace19}. {This factor has been selected to improve the match of observations and simulations at Earth. We have used the same factor in \citet{Singh20b}.} We estimate the radial and azimuthal components of the magnetic field at 0.1 AU from the WSA B values using the local SW speed to account for the Sun's rotation because the SW propagates radially outward from 1 $R_\odot$ to 0.1 AU \citep{MacNeice11}. We reduce the WSA speeds by 20\% to account for the difference in SW acceleration between the WSA and MS-FLUKSS models \citep[e.g.,][]{MacNeice11, Kim14}. {\citet{Kim14} showed that using this reduction factor results in an improved matching of observed and simulated parameters at Earth.} To estimate the SW density and temperature at 0.1 AU, we use the empirical correlations between the SW speed, density, and temperature based on OMNI data \citep{Elliott16}. These procedures give us all inner boundary conditions required to simulate SW in the IH.

\subsubsection{CME model}
\begin{figure}[!htb]
\centering
\center
\begin{tabular}{c c} 
\includegraphics[scale=0.1,angle=0,height=5cm,keepaspectratio]{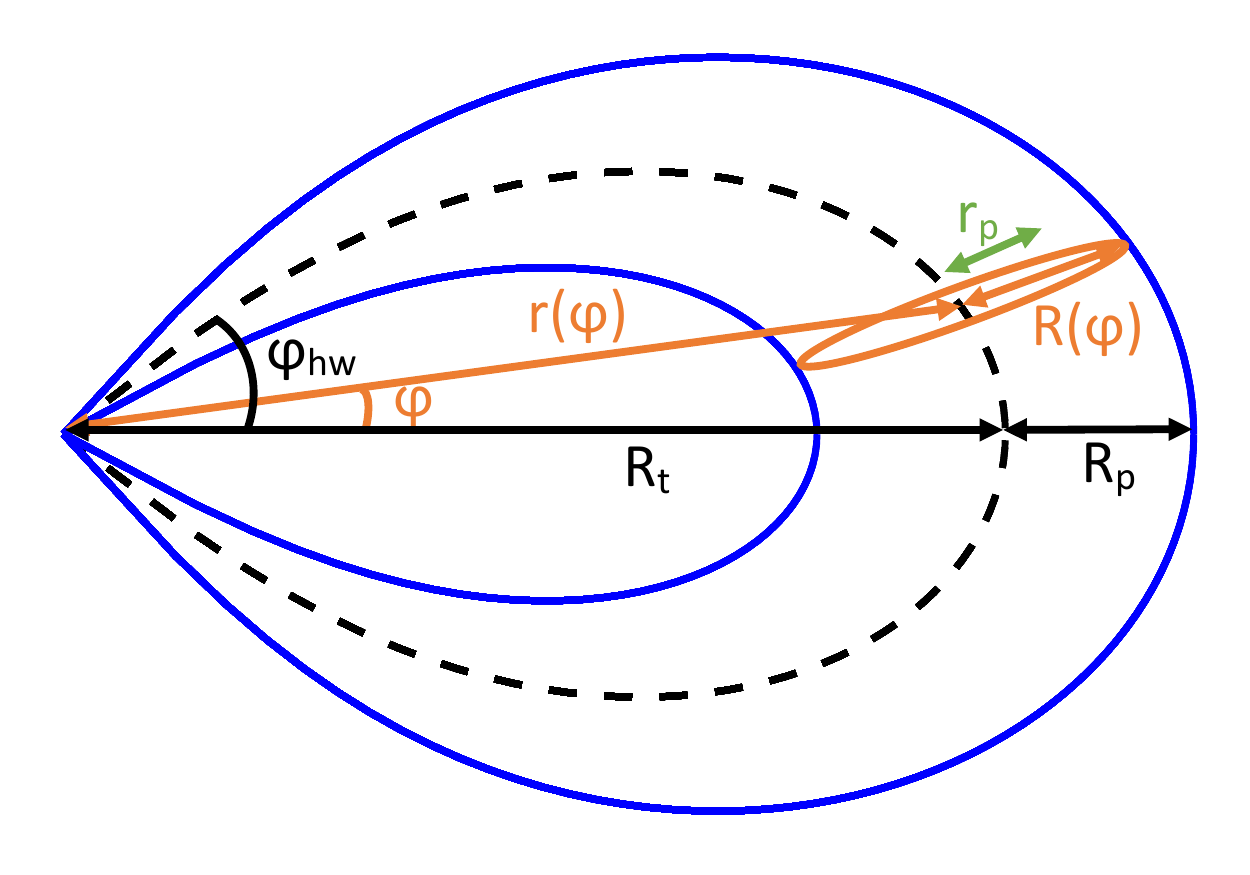} 
\includegraphics[scale=0.1,angle=0,height=5cm,keepaspectratio]{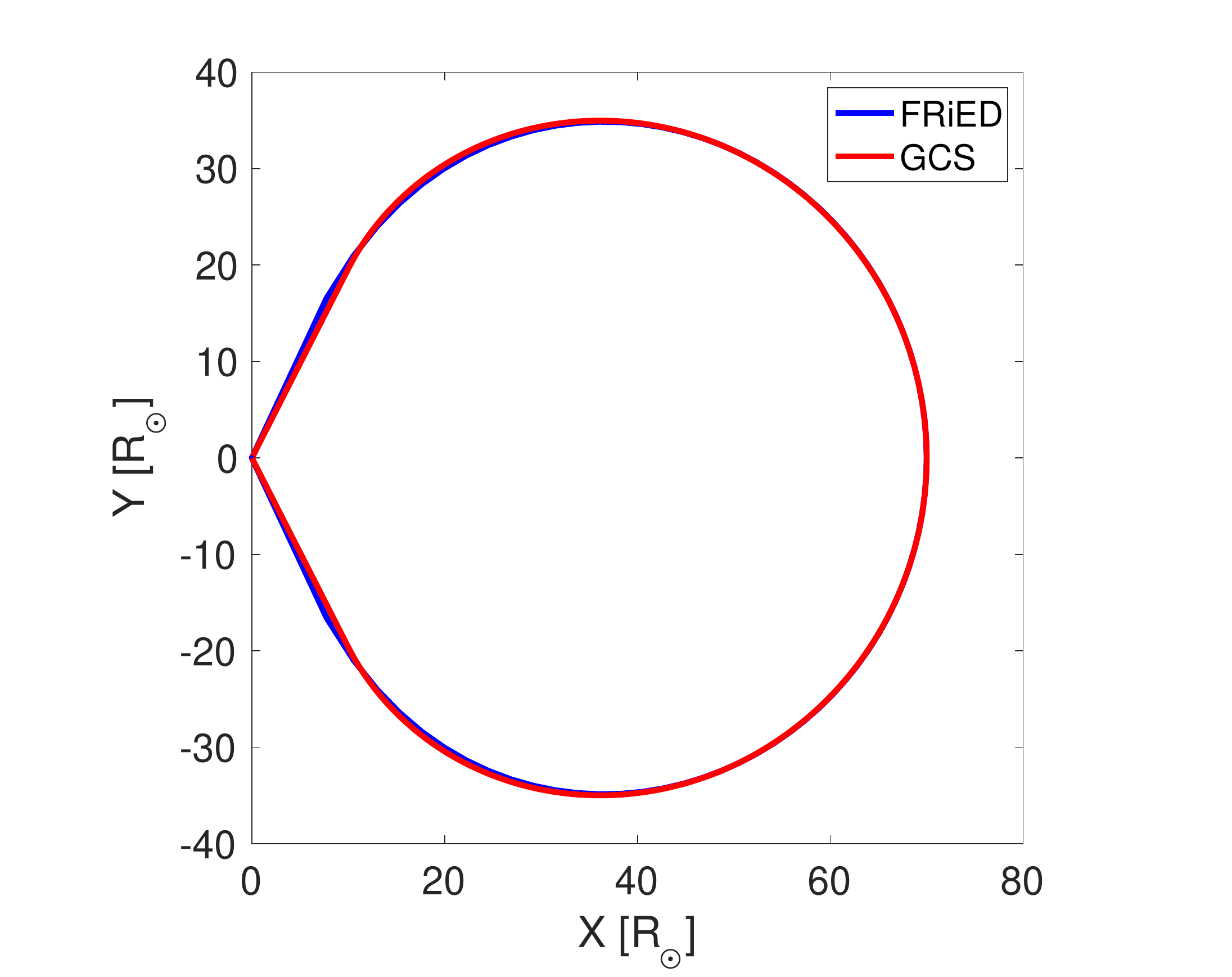} 
\end{tabular}
\caption{(\textit{Left panel}) {A cross-section of constant-turn flux rope by the plane containing the curved axis of this flux rope (shown here by the {dashed} black line). Blue lines represent the inner and outer edges of the flux rope in this plane. The variable $r(\phi)$ is a polar coordinate describing the curved axis. $R(\phi)$ is the radius of the out-of-plane circular cross section (shown with orange color) centered at a point belonging to the curved axis, while $r_p$ is the distance of a point on this circular cross section from its center.} $R(\phi)$ and $r(\phi)$ are given by Equations~\ref{eq:R_phi} and \ref{eq:r_phi} respectively. (\textit{Right panel}) The outer edge of the FRiED model with model parameters $R_\mathrm{t} = 46.67R_\odot$, $R_\mathrm{p} = 23.33R_\odot$, $\phi_\mathrm{hw} = 40^\circ$, and $n=0.33$, is compared with the outer edge of the GCS model with model parameters $\textup{half width} = 33^\circ$, $\textup{height} = 70R_\odot$ and  $\textup{aspect ratio} = 0.5$.}
\label{GCS_vs_FRiED}
\end{figure}

In this work, to simulate flux-rope-based CMEs, we use the geometry of the FRiED model \citep{Isavnin16} and describe the initial magnetic field in it in accordance with the uniform-twist analytic solution described by \citet{Vandas17}. The geometry of the FRiED model simplifies the CME shape to a croissant-like structure with two legs rooted at the center of the Sun, as shown in the left panel of Figure~\ref{GCS_vs_FRiED}. This geometry is characterized by the parameters $R_\mathrm{t}$, $R_\mathrm{p}$, and $\phi_\mathrm{hw}$, as shown in the figure. The flux rope has a circular cross-section with a curved axis shown by the dashed black line. The radius of this cross-section, $R(\phi)$, is specified using the formula
\begin{equation}\label{eq:R_phi}
    R(\phi) = \frac{R_\mathrm{p}}{R_\mathrm{t}}r(\phi),
\end{equation}
where $r(\phi)$ is the distance of the curved axis from the origin. 

This distance is defined by the formula
\begin{equation}\label{eq:r_phi}
    r(\phi) = R_\mathrm{t}\textup{cos}^n\left(\frac{\pi}{2}\frac{\phi}{\phi_\mathrm{hw}}\right),
\end{equation}
where, $n$ is a free parameter which can be used to adjust the flatness of the flux rope shape.

We found that the flux rope geometry of the FRiED model can be matched very well with the GCS model, designed with any suitable choice of its parameters, by appropriately selecting the FRiED parameters. For example, the outer boundary of a GCS-shaped flux rope with half-width $hw_\mathrm{GCS} = 33^\circ$, height $H_\mathrm{GCS} = 70\,R_\odot$, and aspect ratio $\kappa = 0.5$ matches very well with the FRiED flux rope with $R_\mathrm{t} = 46.67\,R_\odot$, $R_\mathrm{p} = 23.33\,R_\odot$, half-width $\phi_\mathrm{hw} = 40^\circ$, and $n=0.33$, as shown in the right panel of Figure~\ref{GCS_vs_FRiED}. The choice of parameters for these shapes is suitable for the 12 July 2012 CME, which we are going to study below.  

We use the uniform twist solution of \citet{Vandas17} because some recent studies \citep[e.g.,][]{Hu15} show that in situ signatures of the ICME magnetic field are consistent with a constant turn configuration. Since the FRiED model has a curved geometry, we cannot use constant turn cylindrical flux rope models such as Gold--Hoyle model \citep{Gold60} to describe the magnetic field in it. However, a magnetic field solution in a curved shape such as a torus is more compatible with FRiED model geometry. \citet{Vandas17} give the analytic solution of a constant turn magnetic field in a torus shape as: 
\begin{equation}\label{Eq:B}
B_\mathrm{r} = 0; \\
B_\phi = \frac{B_\mathrm{0}}{1+b^2r^2};\\
B_\theta = -\frac{B_\mathrm{0}R_\mathrm{0}br}{(1+b^2r^2)(R_\mathrm{0}+r\cos\theta)}.
\end{equation}

Here $r$, $\theta$, and $\phi$ are the coordinates of the toroidally curved cylindrical system, which is described in detail in Section 2 of \citet{Vandas17}. The parameters $B_\mathrm{0}$ and  $b$ control the magnetic flux and the number of magnetic field line turns, respectively. $R_\mathrm{0}$ is the major radius of the torus. Since the FRiED geometry is not exactly a torus, we define the magnetic field distribution in the structure by assuming it to be locally toroidal. The parameter $B_\mathrm{0}$ is modified inside the flux rope to conserve the inserted axial (toroidal) flux. The value of $b$ can be estimated from the inserted poloidal flux inside the flux rope. We fix $R_\mathrm{0}$ throughout the flux rope so that the introduced positive and negative poloidal fluxes are within 20\% of each other. This means that the magnetic flux directed into the plane containing the curved axis of the flux rope is within 20\% of the magnetic flux directed out of this plane.

In this model, we define the initial velocity at any location inside the flux rope as a combination of the radial and expansion velocities, i.e., $\vec{V} = \vec{V}_\mathrm{rad} + \vec{V}_\mathrm{exp}$, where $\vec{V}_\mathrm{rad}$ is in the radial direction away from the origin (the Sun's center) and $\vec{V}_\mathrm{exp}$ is in the direction pointing away from the curved axis. The speed at the apex is considered to be the CME speed $V_\mathrm{CME}$. To facilitate a self-similar expansion, we specify
\[
|V_\mathrm{rad}| = \frac{V_\mathrm{CME}}{1+R_\mathrm{p}/R_\mathrm{t}}, \quad |V_\mathrm{exp}(r_\mathrm{p})| = \frac{r_\mathrm{p}}{R_\mathrm{t}}|V_\mathrm{rad}|,
\]
where $r_\mathrm{p}$ is the radial coordinate inside the local torus structure. This specification can be readily derived from the self-similar properties $R_\mathrm{p} \propto R_\mathrm{t}$ and $r_\mathrm{p} \propto R(\phi)$. Initially, we assume a constant density inside the flux rope.

We insert this flux rope into the ambient SW in such a way that the flux rope is initially superimposed with the SW background. This superimposition is described in detail in \citet{Singh20b}. It should be noted that this approach is different from the other, more commonly used approach, where a flux rope is inserted at the inner boundary gradually \citep[e.g.,][]{Shiota16, Verbeke19}. Our method of flux-rope insertion into the SW preserves the CME flux rope much more accurately, exhibiting a curved front and two legs. On the other hand, the gradual, step-by-step insertion of flux ropes at the inner boundary can be problematic because of their unnatural initial expansion, which must then be compensated by introducing the flux rope with initial speeds smaller than observed. We introduce the flux rope parameters as follows:
\begin{itemize}
    \item $\vec{B}_\mathrm{final} = \vec{B}_\mathrm{FR}$,
    \item $\rho_\mathrm{final} = \rho_\mathrm{FR} + \rho_\mathrm{SW}$,
    \item $e_\mathrm{final} = e_\mathrm{FR} + e_\mathrm{SW}$.
\end{itemize}

Here $\vec{{B}}_\mathrm{FR}$ is given by Eq. (\ref{Eq:B}), $\rho$ is the plasma density and $e$ is the total energy density. We also note that
\[
e_\mathrm{FR} = \frac{|\vec{B}_\mathrm{FR}|^2}{8\pi}, \quad e_\mathrm{SW} = \frac{p_\mathrm{SW}}{\gamma - 1} + \frac{|\vec{B}_\mathrm{SW}|^2}{8\pi} + \frac{\rho_\mathrm{SW} |\vec{v}_\mathrm{SW}|^2}{2},
\]
where $p$ and $\vec{v}$ are the thermal pressure and bulk velocity, respectively. We have kept the adiabatic index $\gamma=1.5$ in this study.

\begin{figure}[!htb]
\centering
\center
\begin{tabular}{c c} 
\includegraphics[scale=0.1,angle=0,height=6.5cm,keepaspectratio]{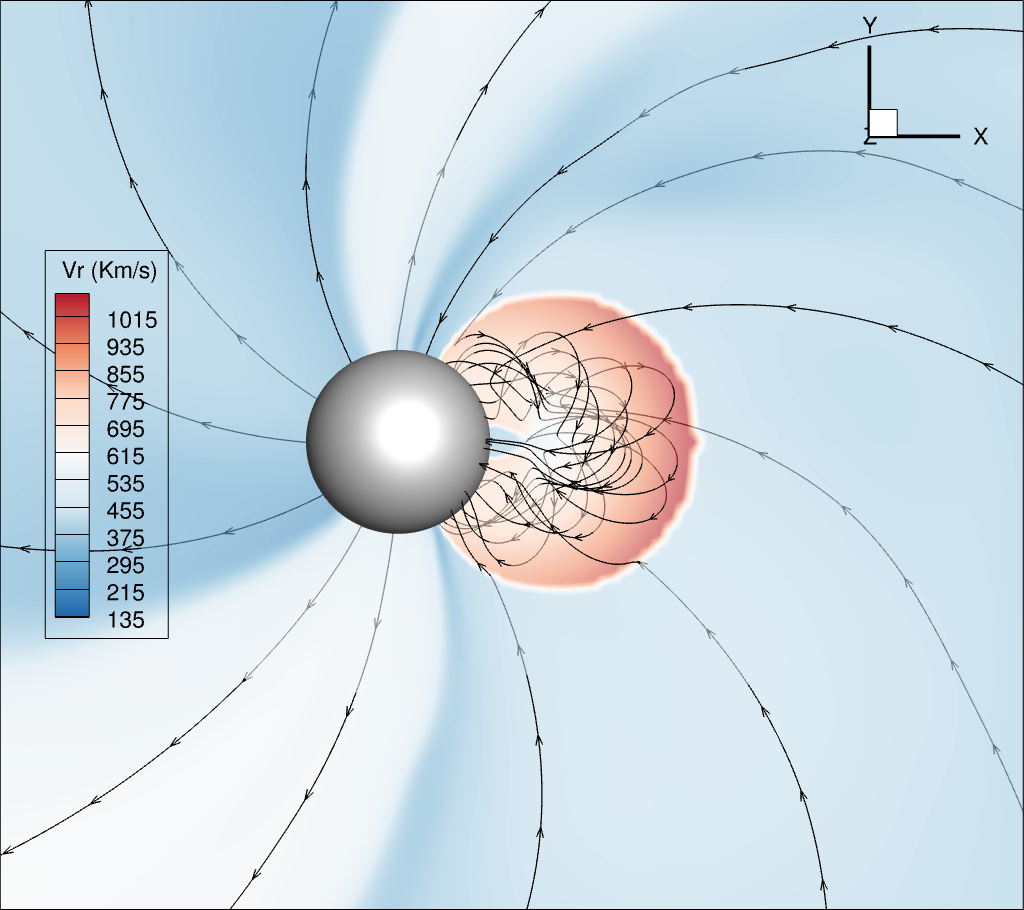} 
\includegraphics[scale=0.1,angle=0,height=6.5cm,keepaspectratio]{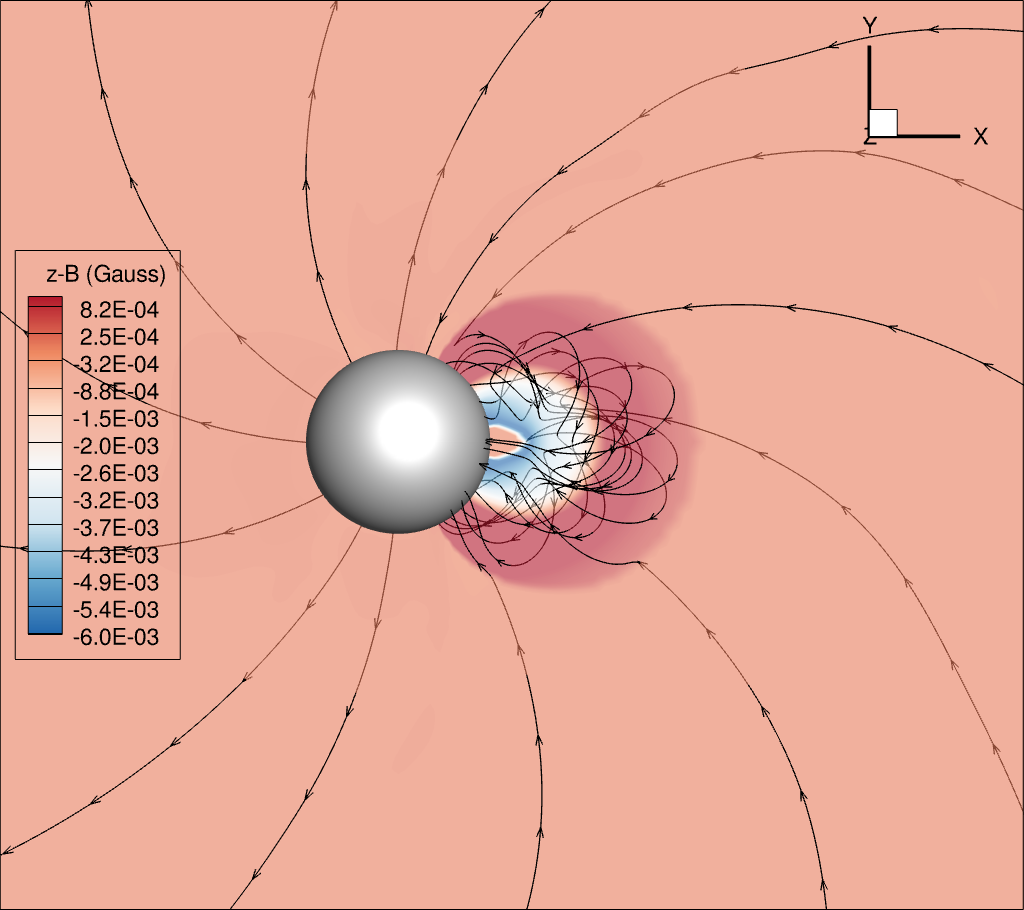} 
\end{tabular}
\caption{The initial structure of the constant turn flux rope model with the apex speed of 1100 km/s, poloidal flux of $1.4\times10^{22}$ Mx, and toroidal flux of $7.6\times10^{21}$ Mx when inserted into the inner heliosphere. The apex of the flux rope is at $70R_\odot$, while its latitude, longitude, and tilt zero. The $z=0$ semi-translucent slices of (\textit{left panel}) radial velocity and (\textit{right panel}) $B_\mathrm{z}$ are shown along with the magnetic field lines. The gray sphere represents the inner boundary of the IH model at 0.1 AU.}
\label{Initial_FR}
\end{figure}

In Figure~\ref{Initial_FR}, we show an example of such insertion. The flux rope has the following FRiED parameters:  $R_\mathrm{t} = 46.67$, $R_\mathrm{p} = 23.33$, $\phi_\mathrm{hw} = 40^\circ$, and $n = 0.33$. The flux rope is assigned an initial speed of 1100 $km/s$, a poloidal flux of $1.4\times10^{22}$ Mx, and a toroidal flux of $7.6\times10^{21}$ Mx. It is initialized with zero latitude, longitude, and tilt. The helicity sign is kept positive. {A flux rope with these properties is chosen because we will later simulate the 12 July 2012 CME with the same parameters.} The left panel of Figure~\ref{Initial_FR} shows the velocity distribution in the $z=0$ slice of this flux rope, along with the magnetic field lines inside it. The right panel of Figure~\ref{Initial_FR} shows the distribution of $B_\mathrm{z}$ in the $z=0$ plane. The gray sphere represents the inner boundary of our simulation domain at 0.1 AU. The SW background shown here is on 12 July 2012 at 21:30 UT.

\begin{figure}[!htb]
\centering
\center
\begin{tabular}{c c} 
\includegraphics[scale=0.1,angle=0,height=6.5cm,keepaspectratio]{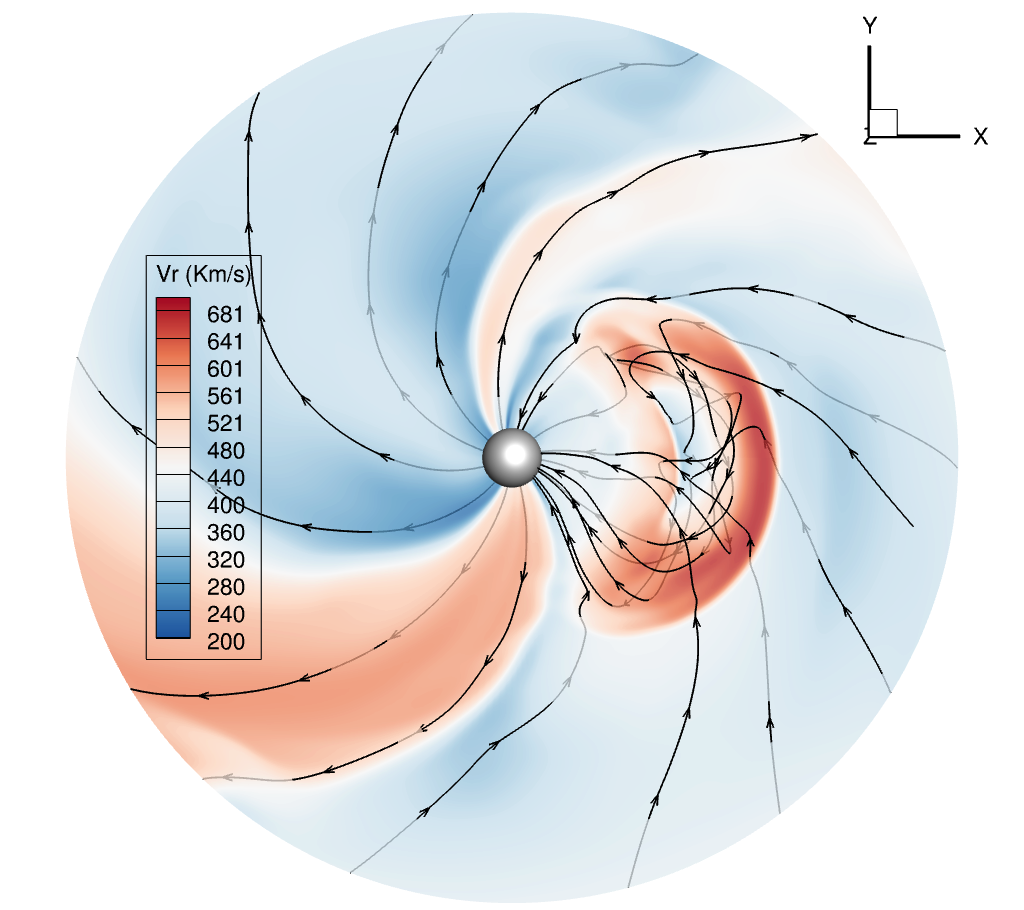} 
\includegraphics[scale=0.1,angle=0,height=6.5cm,keepaspectratio]{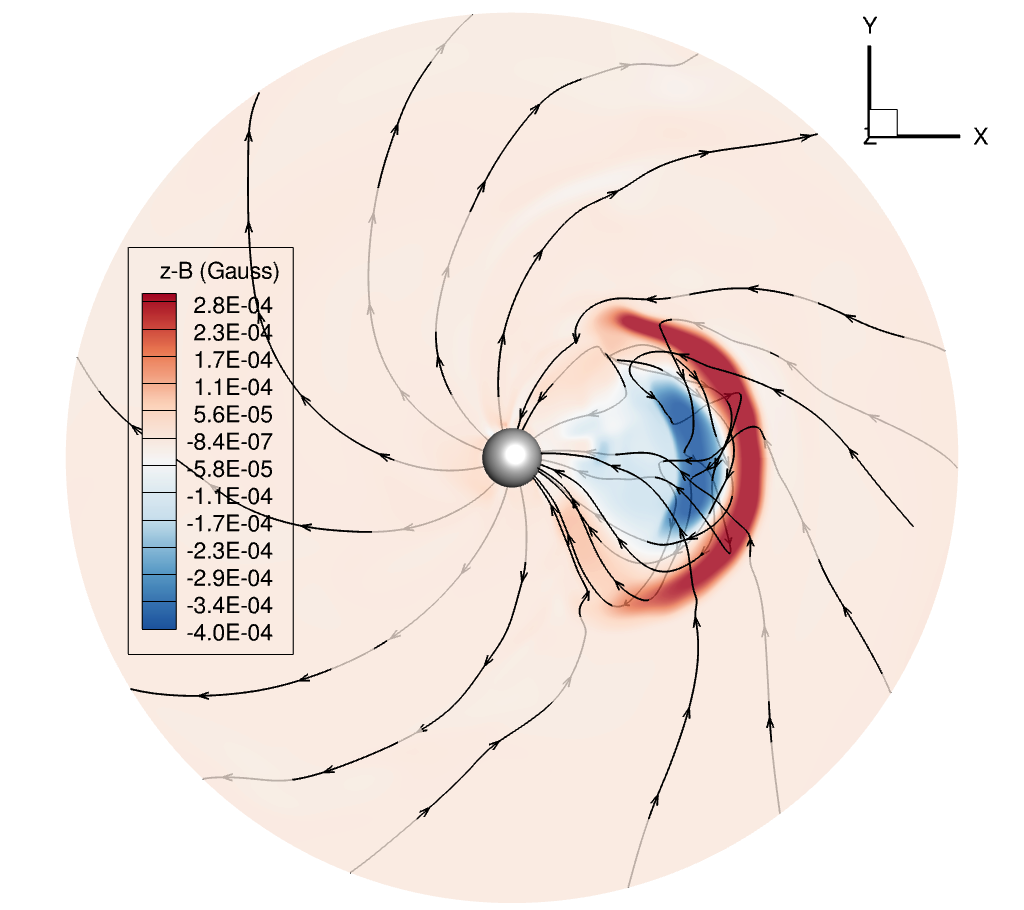} 
\end{tabular}
\caption{The flux rope shown in Figure~\ref{Initial_FR} 30 hours after the initial insertion is propagating as an ICME. The $z=0$ semi-translucent slices of (\textit{left panel}) radial velocity and (\textit{right panel}) $B_\mathrm{z}$ are shown along with the magnetic field lines. The gray sphere represents the inner boundary of the IH model at 0.1 AU.}
\label{Final_FR}
\end{figure}

After the initial insertion, the model flux rope propagates through the IH as an ICME. In Figure~\ref{Final_FR}, We show the model ICME after 30 hrs of propagation. The apex speed of the ICME has dropped to about 700 $km/s$ by this time and the $B_\mathrm{z}$ values have also reduced due to the ICME expansion. This example shows that our model can be used to simulate ICMEs in the IH.

\section{Results}\label{sec:Results}
\subsection{12 July 2012 CME}\label{sec:July12CME}
In this section, we present the simulation results for the 12 July 2012 CME.
We have found the linear speed, latitude, longitude, tilt, half-angle, and aspect ratio of this CME to be 1265 km/s, $-10^\circ$, $1^\circ$, $52^\circ$, $33^\circ$, and 0.5, respectively, at $15\,R_\odot$. Since we introduce our model CME into the SW background in a single step, we have to introduce the CME when its apex has already reached a height of $70\,R_\odot$, as we show in Figure \ref{Initial_FR}. We assumed a self-similar CME expansion from $15R_\odot$ to $70R_\odot$ to propagate the model CME from $15\,R_\odot$ to $70\,R_\odot$. Using the drag-based model (DBM) \citep{Vrsnak07}, we estimated that the CME should reach the apex height of $70R_\odot$ on 13-Jul-2012 03:33 with a speed of about 1097 km/s. We used the drag parameter of $0.1\times10^{-7}$ and the asymptotic SW speed equal to 450 km/s here. {\citet{Vrsnak14} found that this combination of drag parameters ensures roughly the same arrival time accuracy for the DBM and WSA-ENLIL-Cone models.} We also found that the poloidal flux carried by this CME is $1.4\times10^{22}$ Mx. {This value is found from the reconnected flux under the PEA of the source active region using the method described by \citet{Gopalswamy18}}. We used this value in the empirical relation between the ICME poloidal and toroidal fluxes \citep{Qiu07} to calculate the toroidal flux of this CME to be $7.6\times10^{21}$ Mx. The helicity sign of this CME was found to be positive \citep{Singh19}. Using the CME brightness in the coronagraph images, we found that the CME mass is $1.65\times 10^{16}$ g. The brightness is due to the Thomson scattering of photospheric light by the plasma electrons \citep{Billings66}. By integrating over the CME area and removing the projection effects with the help of multiple coronagraph images, we can calculate the true mass of the CME \citep{Colaninno09}, which includes the contribution from the CME sheath, cavity, and core. However, we insert only the flux-rope part of a CME into the ambient SW, which resides in the cavity region and has a density much lower than the sheath and core densities \citep{Riley08}. Therefore, we added only $1.65 \times 10^{13}$ g of uniformly distributed extra mass inside the model flux-rope to the ambient (background) SW. As shown below, introducing this fraction of the total mass ensures a reasonable agreement with the observed ICME density at Earth. {However, it needs to be investigated further if this fraction of total mass is appropriate for other CMEs.}

\begin{figure}[!htb]
\centering
\center
\begin{tabular}{c} 
\includegraphics[scale=0.1,angle=0,height=12cm,keepaspectratio]{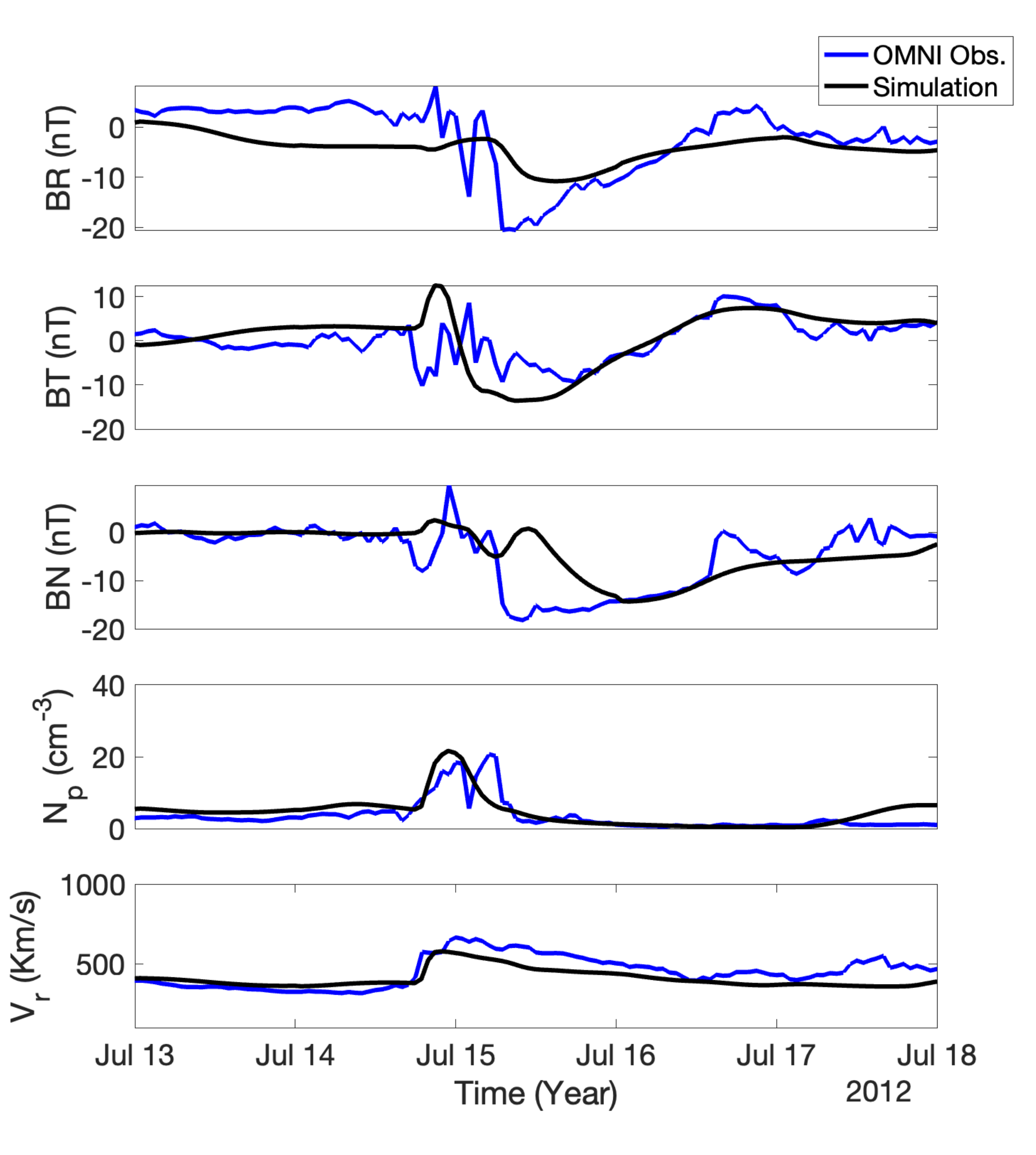} 
\end{tabular}
\caption{Comparison of the 1 hr averaged OMNI SW data (blue) and simulation results at Earth (black). The  simulated  ICME arrived 1 hour after the observed ICME. The signs of the peaks in the magnetic field distribution inside the ICME cloud agree well for all three components. The $N$-component of any vector in the $RTN$ coordinate system is comparable to the $z$-component in the GSE coordinate system.}
\label{July12CME}
\end{figure}

Figure~\ref{July12CME}  shows our comparison between the observations and simulations at Earth. We use the 1 hour averaged SW data provided by NASA/GSFC
through OMNIWeb \citep{King05}. Visual inspection shows that the ICME arrives at Earth with a turbulent sheath, followed by a magnetic cloud with a smoothly rotating magnetic field. The distribution of magnetic field in the cloud has peaks in the negative direction for all $RTN$ components. Please note that by peaks, we mean maximum excursion, and this definition is valid for the rest of the paper. Comparing our simulation results to observations, we find that there is a {reasonable} agreement in the density, speed, and magnetic field values. Our simulations were able to reproduce the signs of the peaks in all three components of the magnetic field inside the ICME cloud. The peak values of the $RTN$ components of the magnetic field inside the observed cloud were (-12,-9,-18) nT. The corresponding values in the simulated ICME cloud were (-11, -14, -14) nT. {This agreement in the magnetic field seems to be better than  in the previously reported simulation studies of the same CME \citep{Shen14, Scolini19, Singh20b}.} The simulated ICME arrived just 1 hour behind the observed ICME. The simulated density, {both in the sheath region and the magnetic cloud,} matches well with the observations, although we inserted only 0.1\% of the observed mass in the initial flux rope. {However, we do notice that the density in the sheath region is much smoother in our simulation as compared to the observation. We also note that the simulated speed at the ICME front agrees with observations. Inside the ICME, our simulation shows  speeds moderately lower than in the observations.} Worth noticing are some differences between the model background SW and OMNI data ahead of the interplanetary shock arrival on 14 July 2012, which can be a possible source of errors in the sheath region. Firstly, the background magnetic field polarity is different in the modeled and observed SW, which contributes to the errors in the simulated sheath magnetic field. Secondly, the model SW is slightly denser and faster than observed. 

In our previous study, we simulated this CME using the modified spheromak model \citep{Singh20b}, which did not reproduce the negative value of $B_\mathrm{r}$. In future studies, we will compare the modified spheromak and constant turn models in more depth to determine whether the constant turn model performs consistently better than the modified spheromak model. This ICME had been modeled previously by \citet{Shen14} and \citet{Scolini19} with the magnetized plasma blob and spheromak models, respectively. The magnetic field components obtained with our constant turn model seem to agree with observations much better than in the above-mentioned studies. 

The kinematic and magnetic properties of CMEs derived from observations contain some level of uncertainty. In the next subsections, we will address the uncertainties associated with GCS fittings and analyze their effects on the CME simulation results at Earth. We only focus on the uncertainties due to GCS fitting. The impact of the uncertainties in the magnetic field flux estimates on the CME properties at Earth will be considered elsewhere.

\subsection{GCS uncertainty analysis}\label{sec:GCS_uncertainity}
The GCS model is a useful tool to derive the 3D properties of CMEs using multiple viewpoint images. However, the derived properties can have large uncertainties, due to the idealization of the CME shape to be a symmetrical, two-conical-legs structure with a curved front that is propagating radially outward. In reality, CMEs can be highly asymmetrical, skewed, and/or slanted with respect to the radial direction. Moreover, SOHO and STEREO A\&B roughly belong to the same ecliptic plane. The direction of the CME out of this ecliptic plane is roughly its latitude value, and it can be easily fitted with the GCS model because all three viewpoints will show a similar extent of the CME out of this plane. The direction of the CME in the ecliptic plane is roughly its longitude value. Estimating this direction with the GCS model is challenging compared to estimating the latitude because all three viewpoints will show different extents of the CME in this direction. This means that GCS fitting based on SOHO and STEREO data can result in larger ambiguities in the longitude estimates as compared to latitude estimates. 

Apparently, the estimates of GCS parameters can vary depending on the fitting approach. Here we try to estimate the average subjective {uncertainties} for the same CMEs when fitted by different GCS users. Using GCS fitting results reported in different publications, as well as in online catalogs, we compiled 6 lists of CMEs. For example, \citet{Gopalswamy14} fitted 77 CMEs that had erupted between Feb 2010 and Jan 2014 and reported their latitudes, longitudes, and speeds. \citet{Thernisien09} reported the latitudes, longitudes, tilts, and half-angles of 26 CMEs that erupted between Nov 2007 and Aug 2008. \citet{Shi15} reported latitudes, longitudes, and speeds of 21 CMEs that erupted between Dec 2008 and Oct 2012. CME Kinematic Database (KINCAT) \citep{Millward13} reports GCS latitudes, longitudes, tilts, speeds, half-angles, and aspect ratios of 122 CMEs that erupted between May 2007 and Oct 2013. The full halo CME (FHCME) catalog \citep{Shen13} has GCS fitting results for the latitudes, longitudes, and speeds of 39 front-side halo CMEs between Dec 2009 and May 2012. Finally, we also used our own unpublished catalog of GCS fittings for 49 CMEs that erupted between Feb 2010 and July 2012. All these GCS fittings apply linear regression to the height-time data to obtain CME speed.

Using the above-mentioned CME lists, we identified the CMEs that appeared in at least two of the lists and published the GCS parameters of those CMEs (see \url{https://doi.org/10.5281/zenodo.5515483}), which allowed us to compare the estimated latitude and longitude values of 56 CMEs, tilt values of 22 CMEs, and speed values of 18 CMEs. We define the subjective uncertainty in a GCS parameter for a CME as the {range of measurements i.e.} difference between maximum and minimum values of this parameter reported in the lists that contained the same CME. For example, if a CME appeared in three of the lists that had the latitude estimates of $1^\circ$, $5^\circ$, and $10^\circ$, the subjective uncertainty in latitude for that CME would be $9^\circ$. The speed uncertainty can be better represented by the speed fractional uncertainty because speed uncertainty itself can be higher for faster CMEs than for slower ones. The speed fractional uncertainty can be specified as
$$\textup{Speed Fractional Uncertainty} = \frac{\textup{Speed Uncertainty}}{\textup{Average speed in separate fits}}$$

\begin{figure}[!htb]
\centering
\center
\begin{tabular}{c c} 
\includegraphics[scale=0.1,angle=0,height=4cm,keepaspectratio]{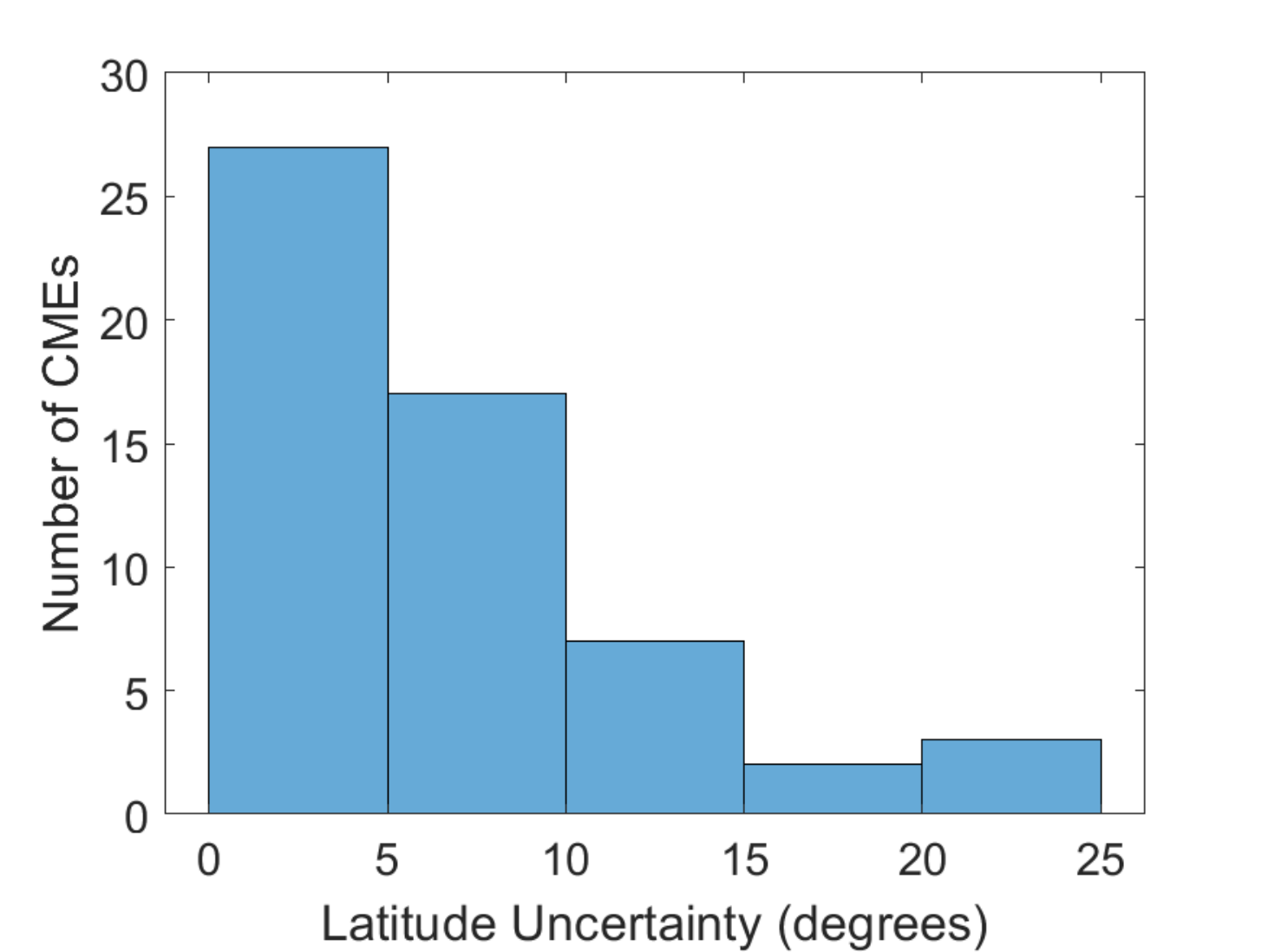} 
\includegraphics[scale=0.1,angle=0,height=4cm,keepaspectratio]{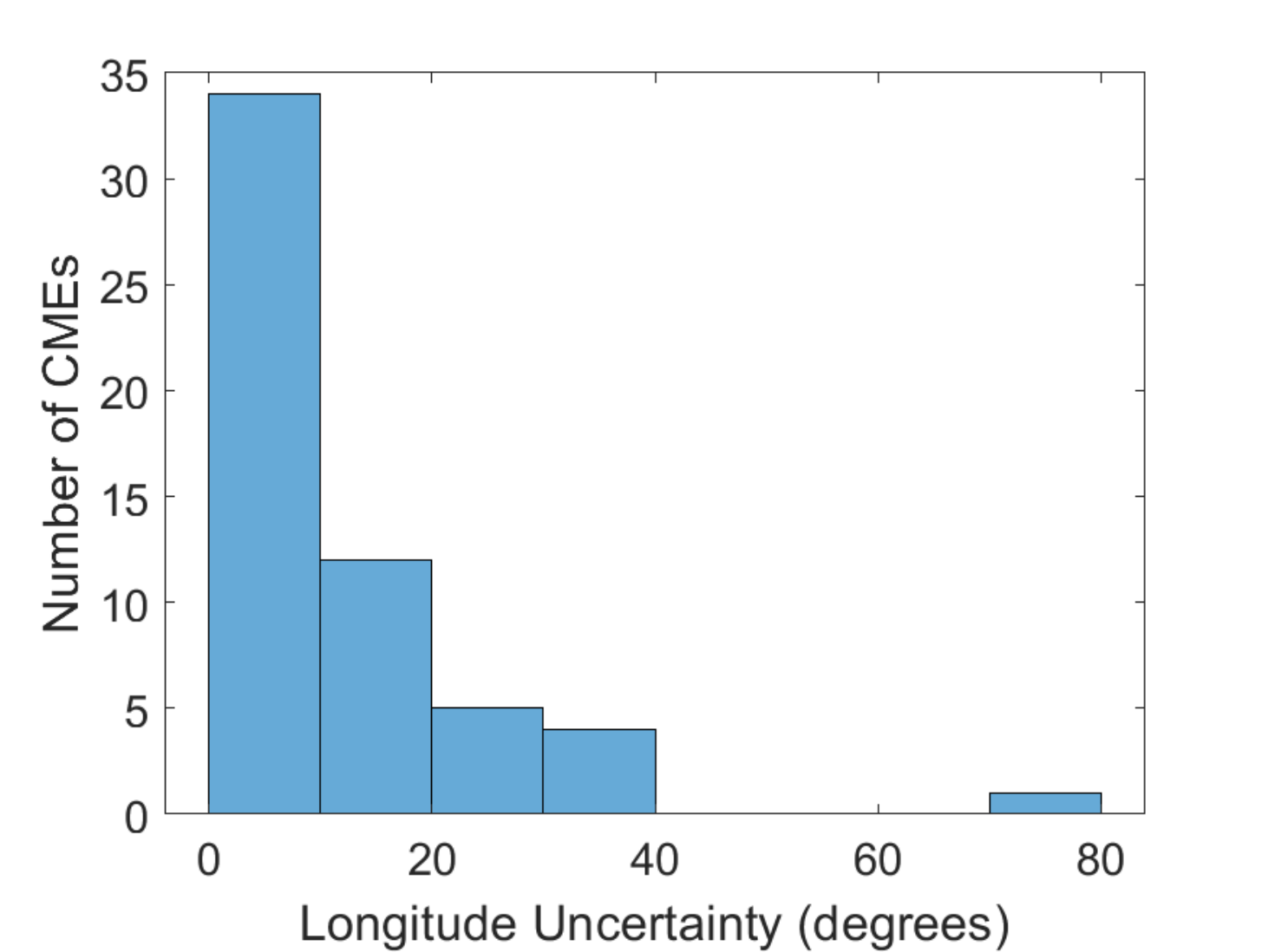} \\
\includegraphics[scale=0.1,angle=0,height=4cm,keepaspectratio]{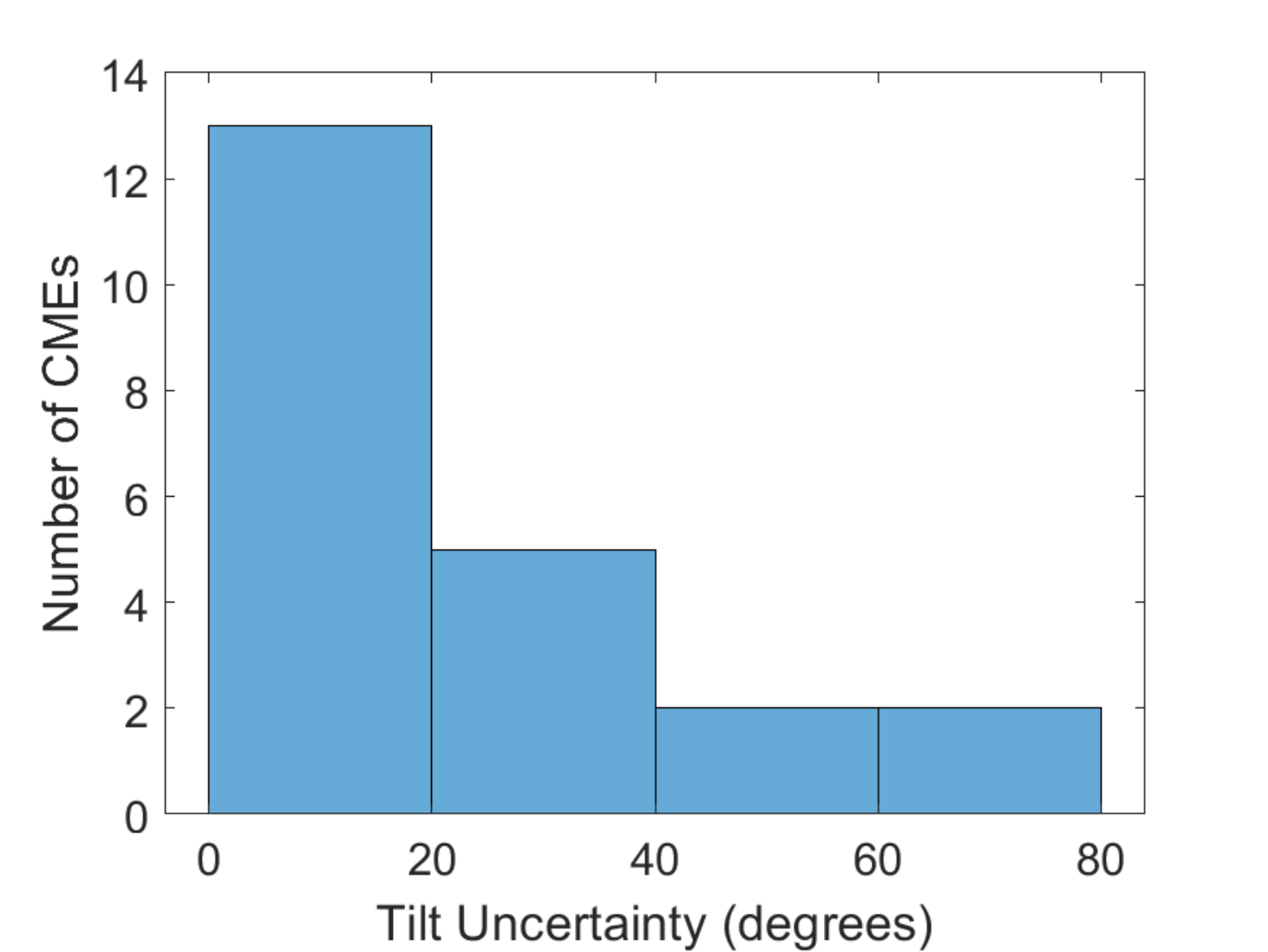}
\includegraphics[scale=0.1,angle=0,height=4cm,keepaspectratio]{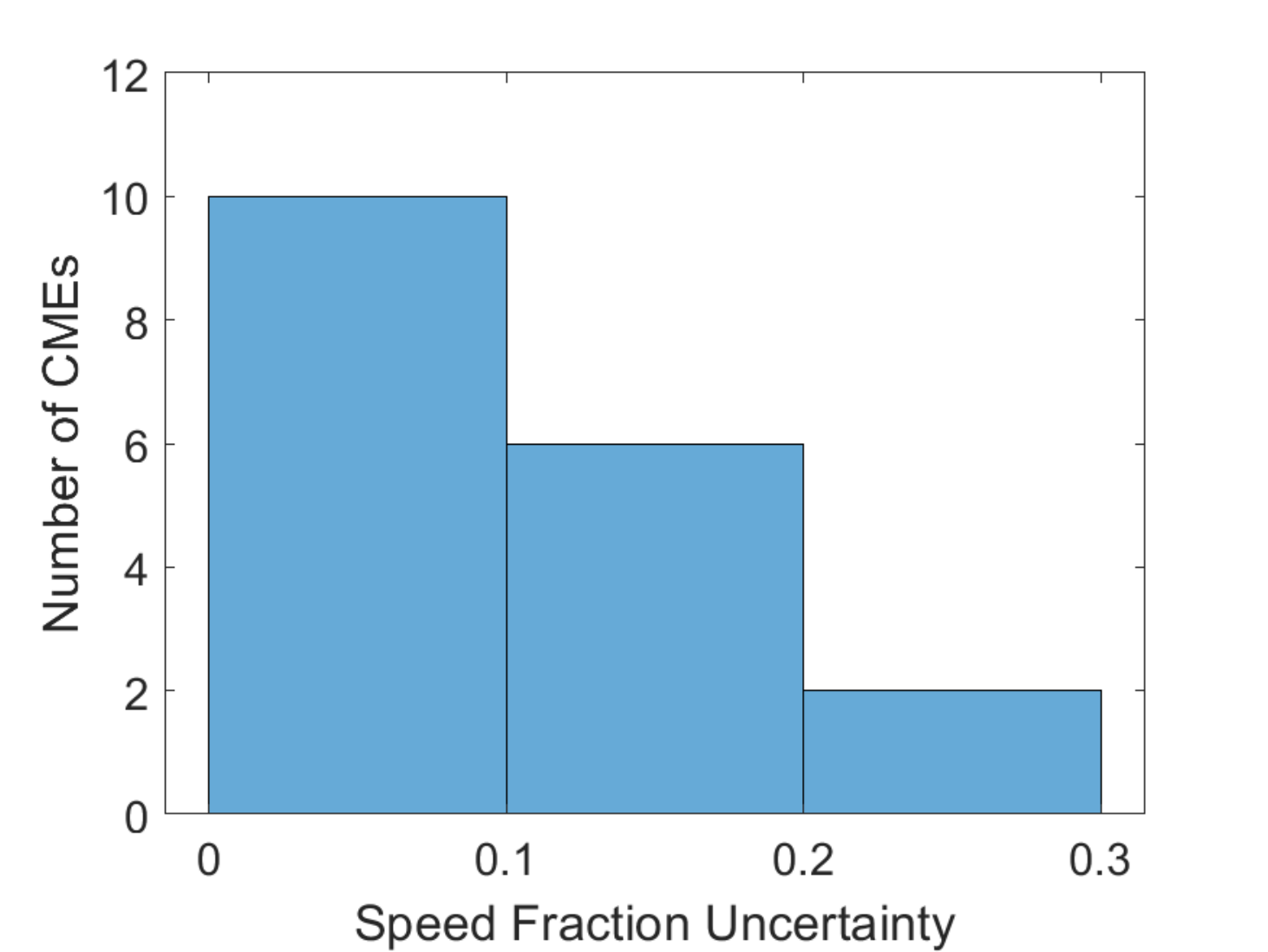} 
\end{tabular}
\caption{Histograms showing the distribution of uncertainties in GCS estimates of (\textit{top-left}) latitude, (\textit{top-right}) longitude, (\textit{bottom-left}) tilt, and (\textit{bottom-right}) speed fraction.}
\label{histograms}
\end{figure}

The distributions of subjective uncertainties for GCS estimates of the considered CMEs are shown in Figure~\ref{histograms}. {We further found the average of these uncertainties.} We found that the GCS estimates of latitude, longitude, and tilt show an average uncertainty of $5.7^\circ$, $11.2^\circ$, and $24.7^\circ$, respectively. We can see that the latitude uncertainties are significantly smaller than the longitude uncertainties, which is because the observing spacecraft are all roughly in the ecliptic plane. The tilt uncertainties are much higher than the latitude and longitude uncertainties. We can see that 4 of the 22 CMEs had tilt uncertainties of more than $40^\circ$. We also found that, on average, CMEs can have  {uncertainties} in their speed of about 11.4\%.

\citet{Thernisien09} used the sensitivity analysis to estimate the average uncertainties in GCS parameters of 26 CMEs they considered in their study. They found the average uncertainties in latitude, longitude, and tilt to be $1.8^\circ$, $4.3^\circ$, and $22^\circ$, respectively, which are smaller than the values we found by comparing the fittings provided by different users of the GCS model. The difference in these values {is} due to the choice of CMEs used in the studies and/or due to the different methods of finding these uncertainties. {\citet{Thernisien09} used a theoretical approach to estimate the uncertainties whereas we used a practical approach in which we directly compared the fitting results of different users of the GCS model.} The average uncertainties in the CME tilt in \citet{Thernisien09} and in our analysis are comparable.

\subsection{Ensemble modeling}\label{ensemble}
\citet{Pizzo15} showed that MHD modeling of CMEs does not depend on the changes in their initial properties chaotically. This means that ensemble modeling of CMEs can be performed to get estimates of the CME forecast uncertainties by creating ensemble members according to the uncertainties in the initial CME model parameters. To study the propagation of uncertainties in the GCS parameters to the CME forecasting uncertainties, we have performed an ensemble modeling for the 12 July 2012 CME using our constant-turn model. We created the ensemble members by taking into consideration the average uncertainties in latitude, longitude, tilt, and speed. Out of the five external lists of reported GCS parameters considered in our study, only the KINCAT database reports half-angles and aspect ratios. Therefore, we could not find the average uncertainties of aspect ratio and half-angle by comparing them across the lists. Instead, we used the uncertainty values reported in \citet{Thernisien09}. They were $+0.07/-0.04$ and $+13^\circ/-7^\circ$ for the aspect ratios and half-angles, respectively. 

We created 77 ensemble members for our CME. These members are described by different GCS parameters used to initiate the CME model. The first member is the one with our own estimated values of the 12 July 2012 CME parameters. An ensemble member with the GCS properties of latitude $\theta$, longitude $\phi$, tilt $\gamma$, speed $V_\mathrm{CME}$, half-angle $\alpha$, and aspect ratio $\kappa$ can be represented by a set $\{\theta, \phi, \gamma, V_\mathrm{CME}, \alpha, \kappa\}$. The other 76 ensemble members can be represented by the following sets:
\begin{enumerate}[noitemsep,nolistsep]
    \item$\{\theta\pm 5.7^\circ, \phi\pm 11.2^\circ, \gamma\pm 24.7^\circ, V_\mathrm{CME}\pm 11.4\%, \alpha_{-7^\circ}^{+13^\circ}, \kappa_{-0.04}^{+0.07} \}$, (64 members)
    \item$\{\theta\pm 5.7^\circ, \phi, \gamma, V_\mathrm{CME}, \alpha, \kappa\}$ (2 members)
    \item$\{\theta, \phi\pm 11.2^\circ, \gamma, V_\mathrm{CME}, \alpha, \kappa\}$ (2 members)
    \item$\{\theta, \phi, \gamma\pm 24.7^\circ, V_\mathrm{CME}, \alpha, \kappa\}$ (2 members)
    \item$\{\theta, \phi, \gamma, V_\mathrm{CME}\pm 11.4\%, \alpha, \kappa\}$ (2 members)
    \item$\{\theta, \phi, \gamma, V_\mathrm{CME}, \alpha_{-7^\circ}^{+13^\circ}, \kappa\}$ (2 members) 
    \item$\{\theta, \phi, \gamma, V_\mathrm{CME}, \alpha, \kappa_{-0.04}^{+0.07}\}$ (2 members)
\end{enumerate}

The 64 members in set 1 represent all possible combinations of GCS parameter extremes, where the GCS parameters we estimated for the 12 July 2012 CME differ according to the average uncertainties in these parameters. The simulation of these ensemble members should give us the upper and lower bounds of simulation {uncertainties} due to subjective uncertainties in the GCS fitting. The 12 members in sets 2--7 are used in this work to study the individual contribution of uncertainties in each parameter to the simulation results at Earth. 

\begin{figure}[!htb]
\centering
\center
\begin{tabular}{c c} 
\includegraphics[scale=0.1,angle=0,height=8.5cm,keepaspectratio]{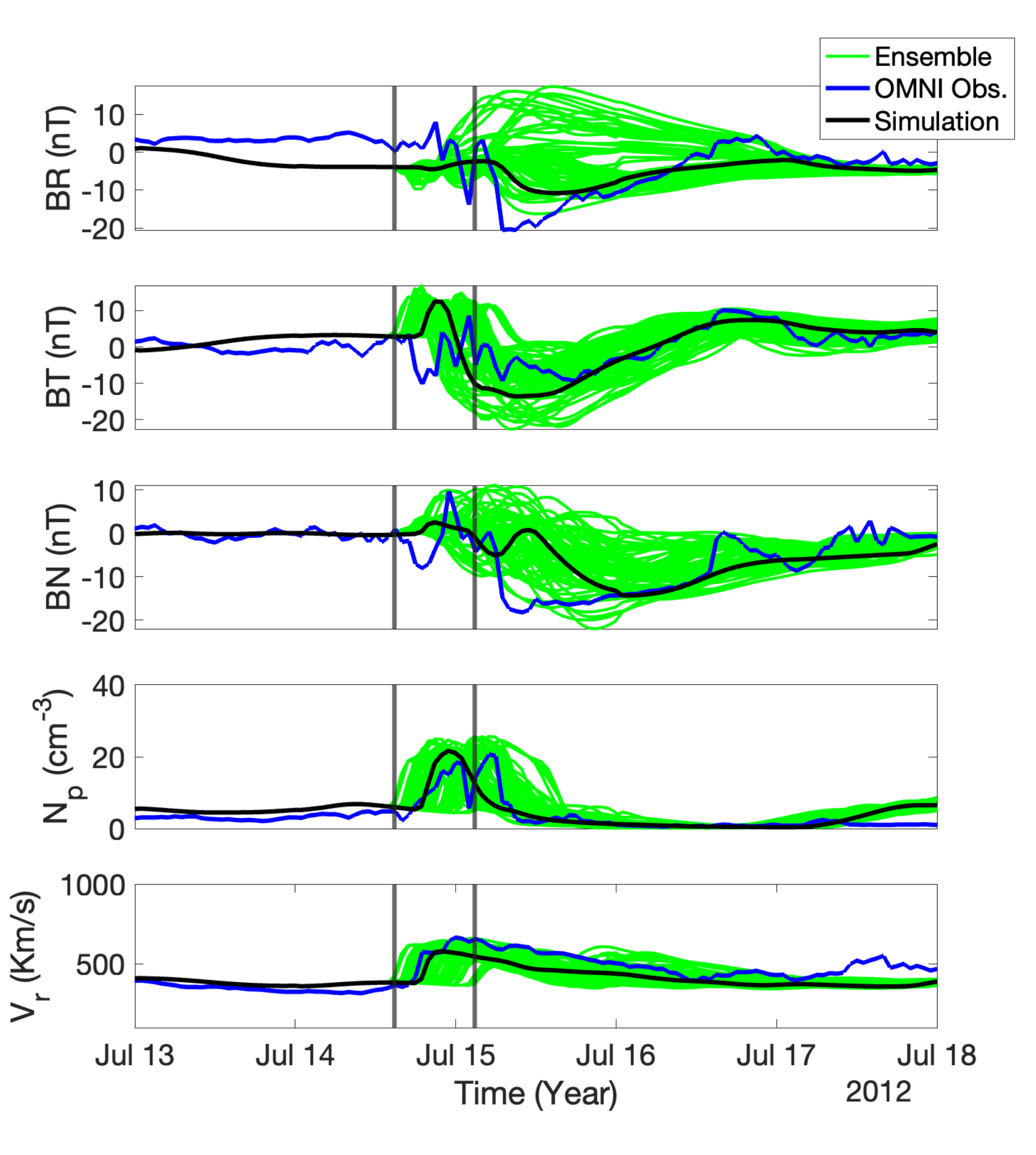} 
\includegraphics[scale=0.1,angle=0,height=8.5cm,keepaspectratio]{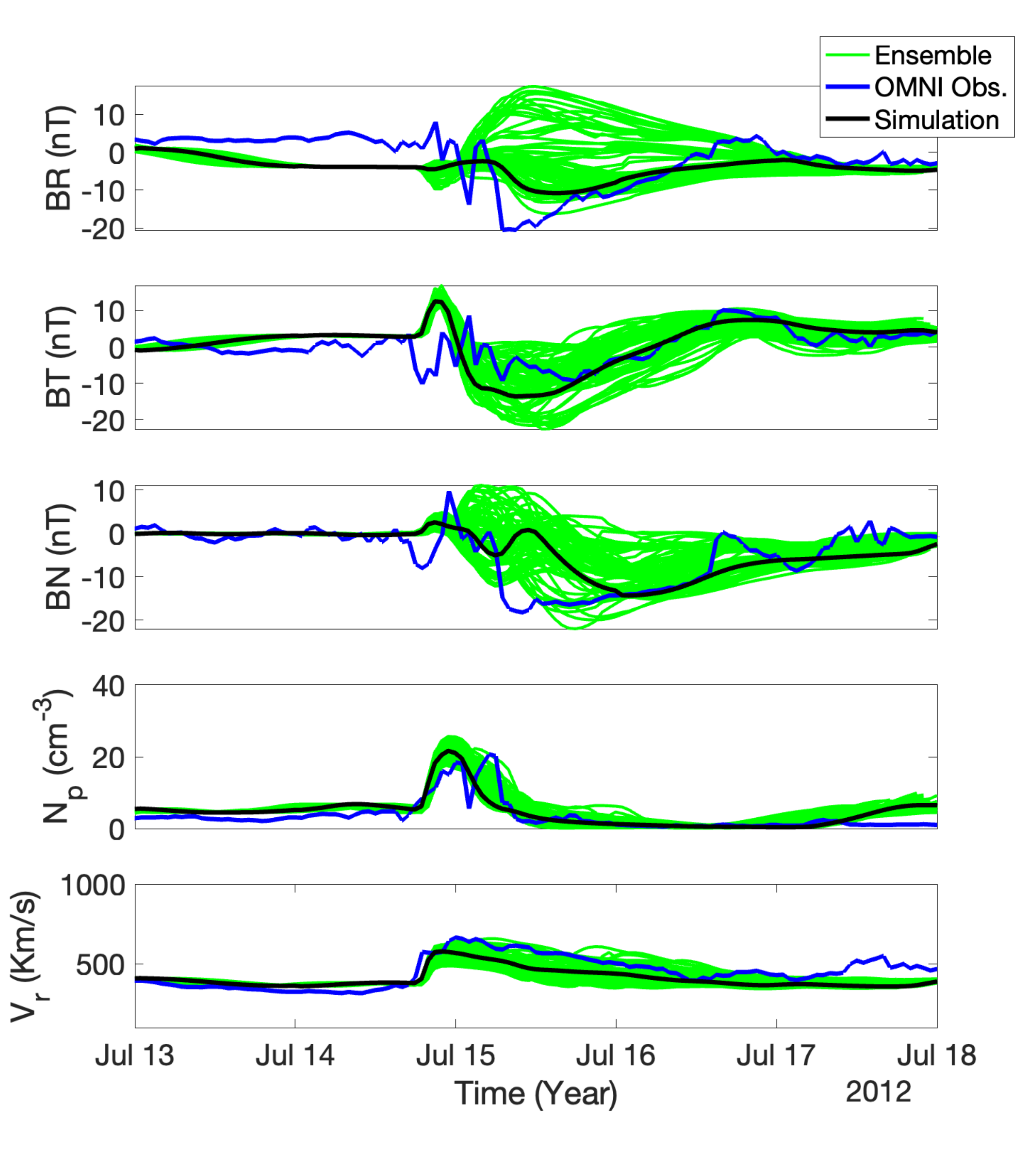}
\end{tabular}
\caption{(\textit{Left panel}) Plasma and magnetic field in the 1 hr averaged OMNI SW data (blue) and simulation results probed at Earth (black and green). The black line represents the first ensemble member with GCS parameters found by us. The green lines represent the rest of the 76 ensemble members defined in Section~\ref{ensemble}. The grey vertical lines show the 12 hour wide arrival time window of all the ensemble members.  (\textit{Right panel}) Same as the left panel, with the arrival time of all ensemble members matched with the first member.}
\label{Full_ensemble}
\end{figure}

In the left panel of Figure~\ref{Full_ensemble}, we compare all ensemble members with OMNI data at Earth. The members show a 12 hr wide arrival window bounded by grey vertical lines. The peak speed at the time of arrival ranges between 480 and 610 km/s. The peak density of the front of ICMEs ranges between 18 and 26  cm$^{-3}$. In the right panel of Figure~\ref{Full_ensemble}, we show the ensemble members by shifting them in time to match the individual arrival times to that of the first ensemble member. This is done for an easier qualitative comparison of the plasma and magnetic field values in the ensemble members. While 55\% of the ensemble members correctly reproduced the peak signs for all three components of the cloud magnetic field, a significant fraction did not correctly reproduce the sign of the peak in the radial magnetic field component.    

To investigate the effects of uncertainty in each GCS parameter on the simulation results, we used sets 2--7, where each of the GCS parameters was modified according to its average uncertainties, keeping the rest of the parameters unchanged. This exercise provides a rather good insight into the role of uncertainties in the individual GCS parameters in producing the CME model uncertainties at Earth. 

\begin{figure}[!htb]
\centering
\center
\begin{tabular}{c c} 
\includegraphics[scale=0.1,angle=0,height=8.5cm,keepaspectratio]{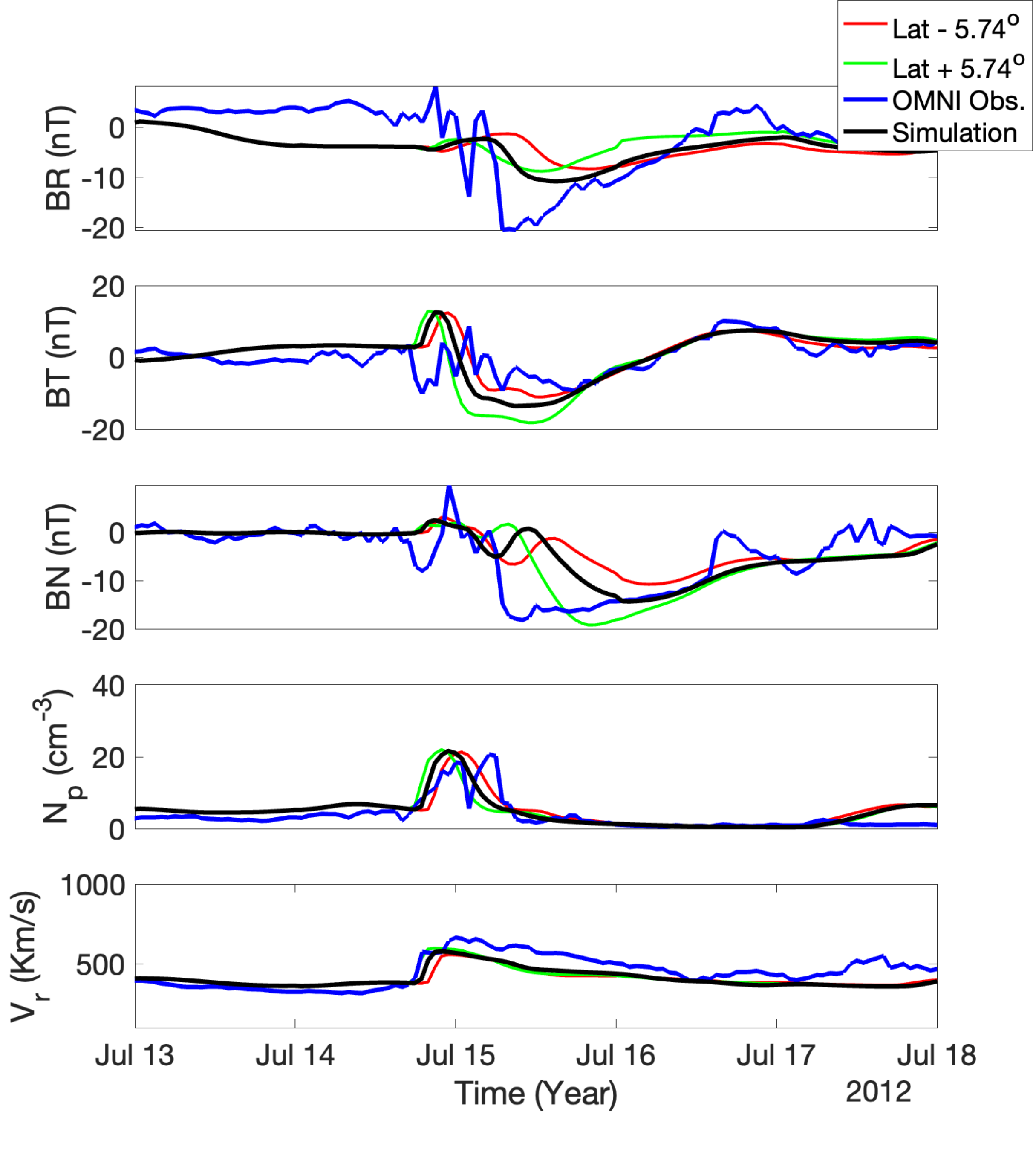} 
\includegraphics[scale=0.1,angle=0,height=8.5cm,keepaspectratio]{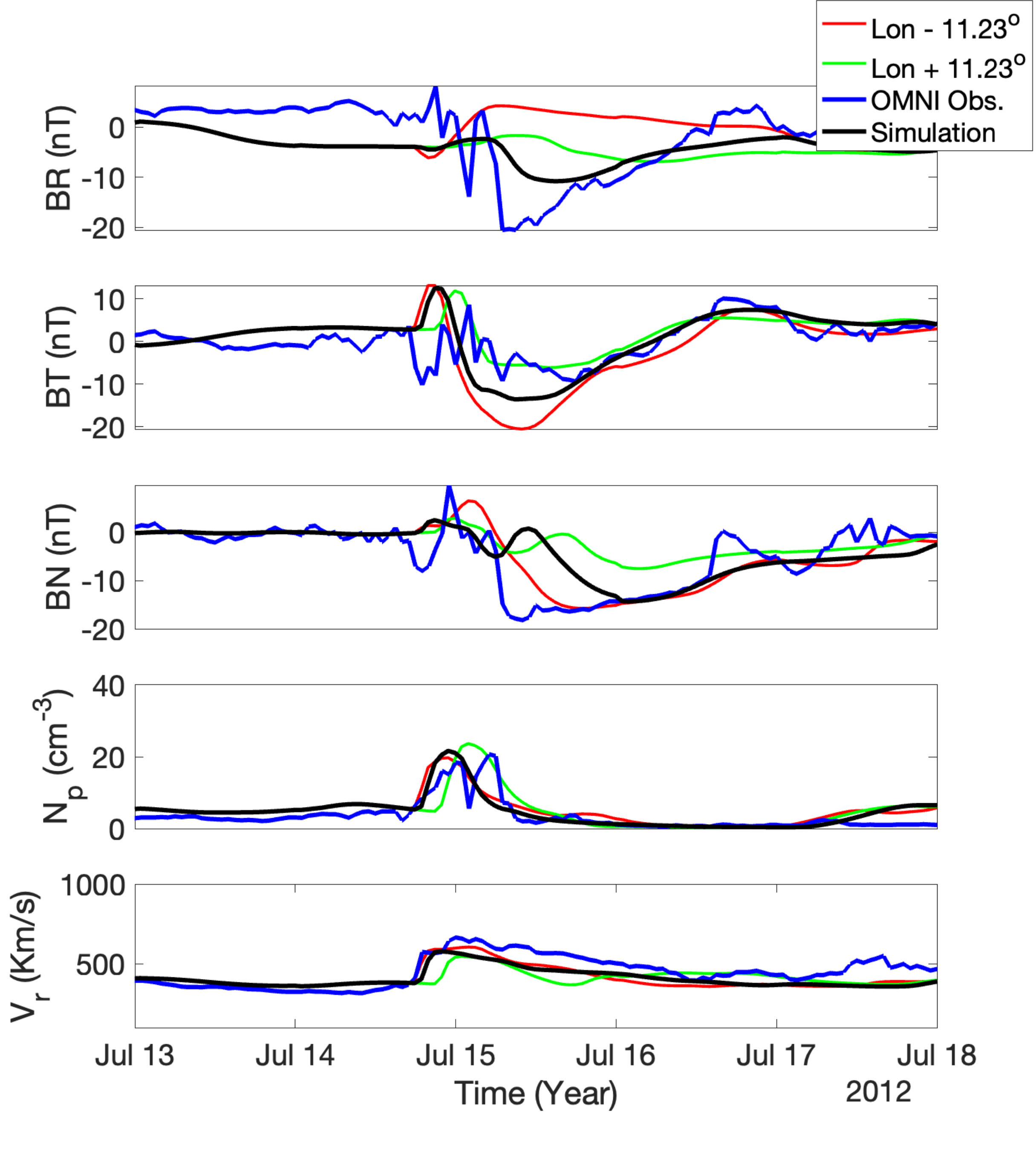} 
\end{tabular}
\caption{The distributions of plasma velocity and density, and magnetic field components in the 1 hr averaged OMNI data (blue) and simulation results at Earth (black, red, and green). The black lines represent the first ensemble member with our derived GCS parameters. The red and green lines represent the two ensemble members defined in set 2 (\textit{Left panel}) and 3 (\textit{Right panel}) in Section~\ref{ensemble}.}
\label{Sim_obs_lat_lon}
\end{figure}

The left panel of Figure~\ref{Sim_obs_lat_lon} shows the effect of uncertainties in latitude on CME forecasting at Earth. The latitude uncertainty of $\pm5.74^\circ$ widens the arrival time window by 2 hours. This {uncertainty}, however, does not change the signs of the peaks in the cloud magnetic field components, which all remain negative. As mentioned in Section~\ref{sec:July12CME}, the peak $RTN$ components of the ICME magnetic cloud of our simulated CME were (-11, -14, -14) nT. By varying the latitude by $\pm5.74^\circ$, we obtain the peak values of the  $RTN$ components equal to (-9, -18, -19) and (-8, -11, -11) nT.  

The effect of the longitude uncertainty on magnetic field forecasts in the cloud is much more pronounced, as can be seen in the right panel of Figure~\ref{Sim_obs_lat_lon}. The longitude uncertainty of $\pm11.23^\circ$ gives the range of the peaks in the $RTN$ components equal to (-7, -7, -7) and (4, -20, -16) nT. We can see that a $-11.23^\circ$ change in latitude results in the wrong sign of  $B_\mathrm{r}$. Besides, a $\pm11.23^\circ$ uncertainty in longitude widens the arrival time window by 3 hours.

\begin{figure}[!htb]
\centering
\center
\begin{tabular}{c c} 
\includegraphics[scale=0.1,angle=0,height=8.5cm,keepaspectratio]{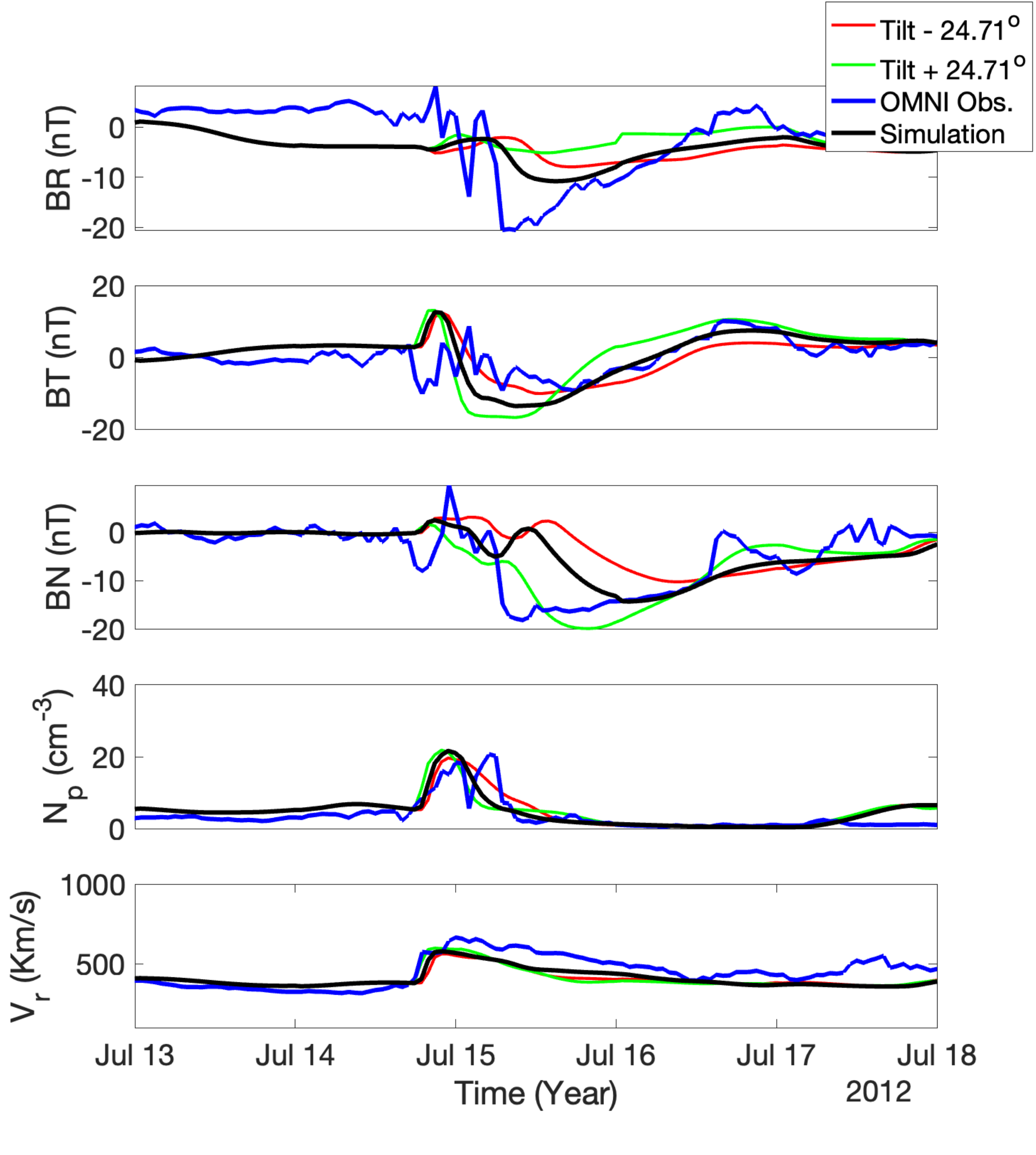} 
\includegraphics[scale=0.1,angle=0,height=8.5cm,keepaspectratio]{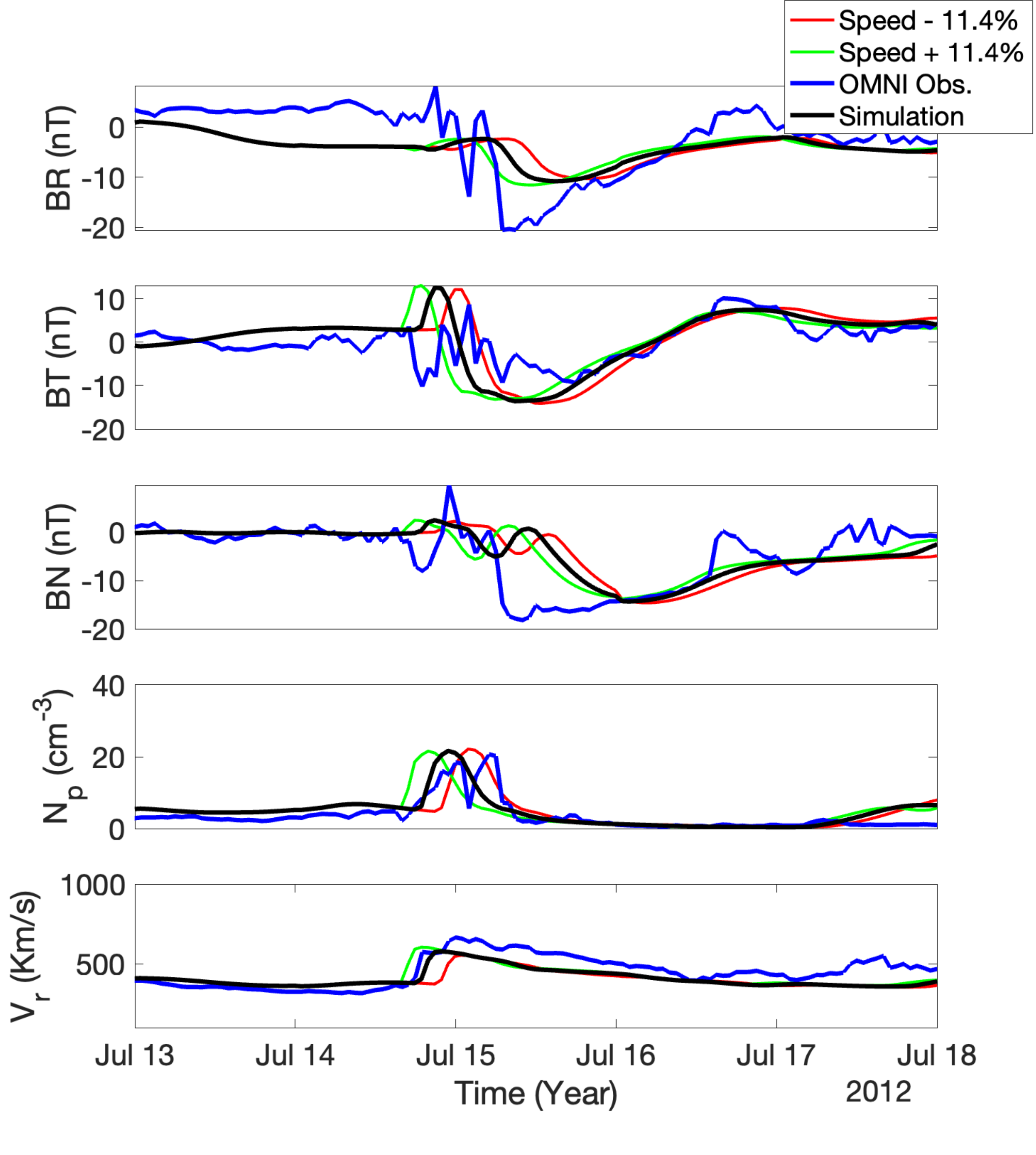} 
\end{tabular}
\caption{Plasma speed and density, and magnetic field component distributions in the 1 hr averaged OMNI SW data (blue) and simulation results at Earth (black, red, and green). The black line represents the first ensemble member with our derived GCS parameters. The red and green lines represent the two ensemble members defined in set 4 (\textit{Left panel}) and 5 (\textit{Right panel}) in Section~\ref{ensemble}.}
\label{Sim_obs_tilt_speed}
\end{figure}

The tilt uncertainty of $\pm24.71^\circ$ can significantly alter the magnetic field profile in the magnetic cloud of ICME at Earth, as shown in the left panel of Figure~\ref{Sim_obs_tilt_speed}. Such tilt range can change the peak values of the $RTN$ components of the cloud magnetic field from (-11, -14, -14) nT to (-5, -17, -20) or (-8, -10, -10) nT. The tilt {uncertainty}, however, does not change the signs of the peak magnetic field components from being negative. The arrival time window is widened by 1 hour due to the tilt uncertainty.

In the right panel of Figure~\ref{Sim_obs_tilt_speed}, we show the effect of a $\pm11.4\%$ uncertainty in speed on the simulated signatures of ICMEs at Earth. One can see that the speed uncertainty does not change the magnetic field profile in the ICME cloud. This uncertainty, however, widens the arrival time window by 6 hours. We also found that the peak density stays unaffected by the ICME speed change. This result is in line with the recent observational analysis by \citet{Temmer21}. It should also be noted that the arrival time window with the full set of ensemble members had a width of 12 hours. This shows that the arrival time window can be widened by as much as 100\%  by the combination of uncertainties in CME parameters, as compared to uncertainties in initial speed only.

\begin{figure}[!htb]
\centering
\center
\begin{tabular}{c c} 
\includegraphics[scale=0.1,angle=0,height=8.5cm,keepaspectratio]{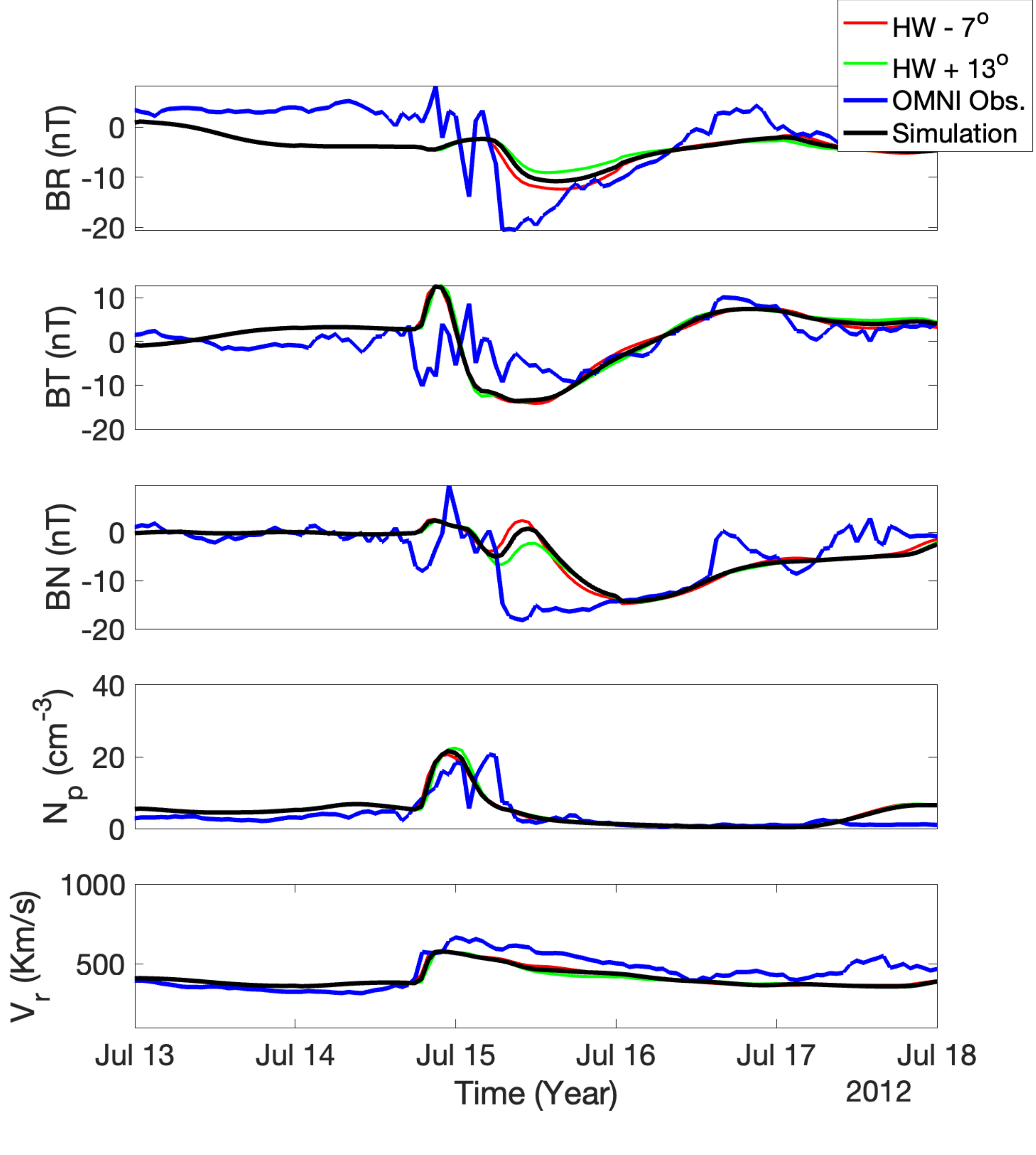} 
\includegraphics[scale=0.1,angle=0,height=8.5cm,keepaspectratio]{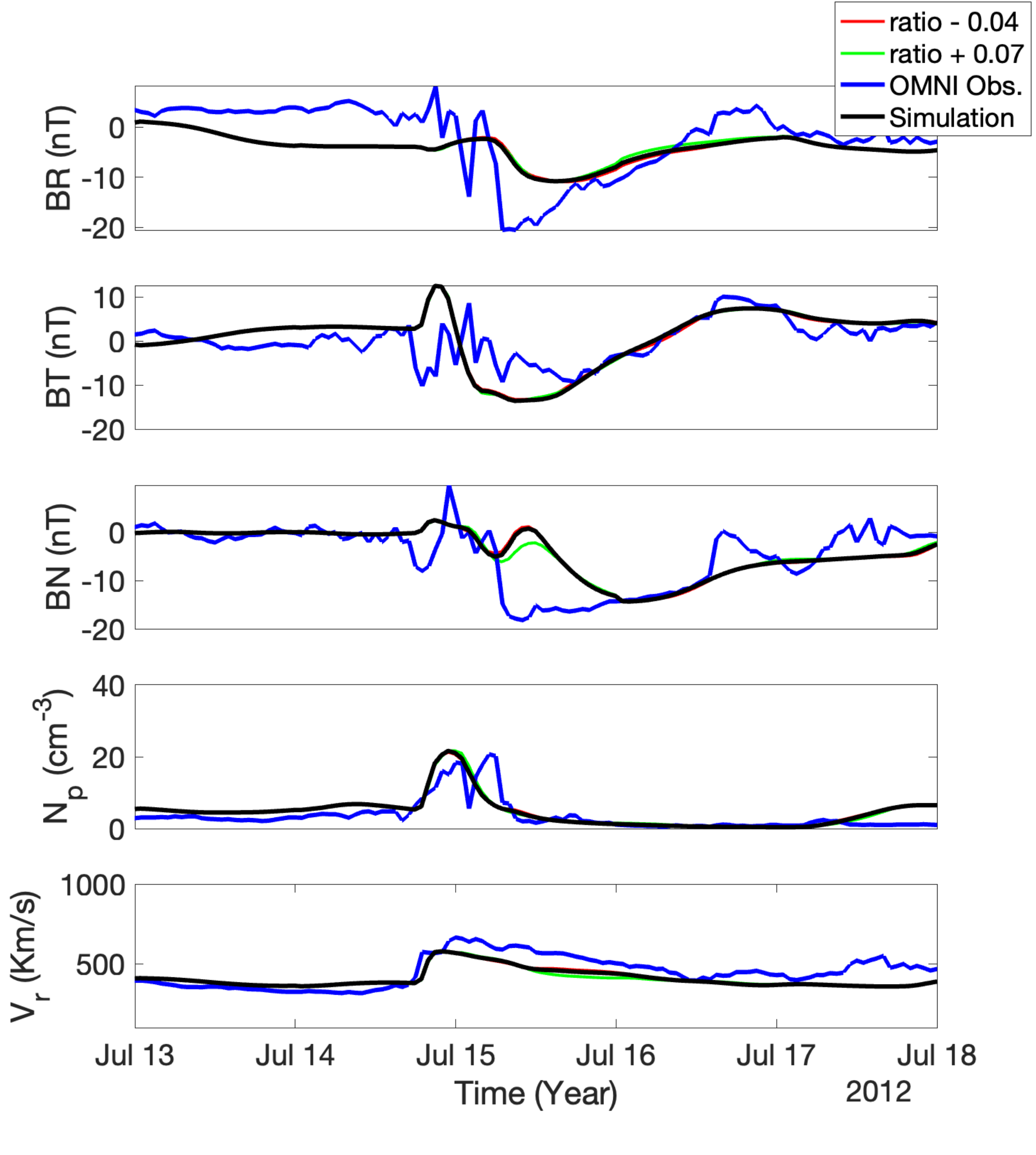} 
\end{tabular}
\caption{Plasma speed and density, and magnetic field components in the 1 hr averaged OMNI SW data (blue) and simulation results at Earth (black, red, and green). The black line represents the first ensemble member with our derived GCS parameters. The red and green lines represent the two ensemble members defined in set 6 (\textit{Left panel}) and 7 (\textit{Right panel}) in Section~\ref{ensemble}.}
\label{Sim_obs_HW_ratio}
\end{figure}

Figure~\ref{Sim_obs_HW_ratio} shows the effect of $+13^\circ/-7^\circ$ half-width $+0.07/-0.04$ aspect ratio uncertainties (the left and right panels, respectively). Both of these uncertainties result in very small changes in the magnetic field and plasma properties at Earth. The arrival time is also largely unaffected by these uncertainties. The ensemble member with the larger half-width has a slightly larger density peak of the front (by 2 cm$^{-3}$), which might be due to a larger mass pile-up at the front during its interplanetary travel. This has been shown also in the observations of \citet{Temmer21}, which show that wider ICMEs tend to have stronger mass pile-up at the front. Since the 12 July, 2012 CME was Earth-directed, we did not expect the half-width to play a major role in controlling the 1 AU signature of this ICME. However, off the Sun-Earth line CMEs can be affected by the uncertainties in half-width to a larger extent. This is because such uncertainties can create an ICME with magnetic clouds either passing over Earth or missing it.

\section{Conclusion}\label{sec:Conclusions}
In this study, we have implemented a flux rope model with a constant-turn magnetic field \citep{Vandas17} to perform MHD simulations of the 12 July 2012 CME propagation through a time-dependent, data-driven ambient SW. The CME shape used in our model is based on the FRiED approach described by \citet{Isavnin16}. It can be matched well with the shape approximation used in the GCS model, which is frequently used to derive 3D properties of CMEs from the analysis of multi-viewpoint images made by coronagraphs. We constrain the poloidal magnetic flux in this model with the measured reconnected flux in the PEAs of the source active region. The toroidal magnetic flux is estimated from the poloidal flux using the empirical relation given by \citet{Qiu07}. We have shown that by constraining the flux rope model with GCS and PEA observations of the 12 July 2012 CME, {the ICME features at Earth are reproduced with reasonable accuracy.} Our simulated CME agrees with the observations of speed, density, and magnetic field components signs in the ICME magnetic cloud. The arrival time of our simulated ICME was off by just 1 hour.

We compared the GCS parameters calculated independently by multiple users to derive the average subjective uncertainties in the CME latitude, longitude, tilt, and speed, which turned out to be  $5.74^\circ$, $11.23^\circ$, $24.71^\circ$, and 11.4\% respectively. Using these uncertainties, as well as those in the CME half-angle and aspect ratio ($+13^\circ/-7^\circ$ and $+0.07/-0.04$, respectively) from \citet{Thernisien09}, we have created 77 ensemble members for the 12 July 2012 CME. The results of our ensemble modeling are summarized as follows:
\begin{enumerate}[noitemsep,nolistsep]
    \item The arrival time window for the full ensemble is 12 hours wide. This value is similar to the typical arrival time errors obtained with different models. This 12-hour width of the full ensemble is due to the combined non-linear impact of uncertainties in all GCS model parameters. 
    \item 55\% of the ensemble members correctly reproduced the signs of all $RTN$ components of the peak magnetic field in the ICME magnetic cloud, while {28\% of} the ensemble members did not produce the correct sign of $B_\mathrm{r}$.
    \item Subjective uncertainties of $\pm11.23^\circ$ in longitude and $\pm24.71^\circ$ in the CME tilt are found to be the main sources of uncertainty in the simulated magnetic field. Subjective uncertainty of $\pm5.74^\circ$ in latitude had a lesser impact on the simulated magnetic field values. This highlights the importance of out-of-ecliptic coronagraph observations of CMEs, which would be very important for better constraining the CME longitude.
    \item Uncertainties in the half-width and the aspect ratio are hardly affecting the ICME features at 1 AU. This is not surprising, since our CME was Earth-directed.  The half-angle of a model CME may play an important role for the CMEs that are not aligned with the Sun-Earth line.
\end{enumerate}
It takes about 2 hours of computation for MS-FLUKSS to simulate about 6 days of physical time on a $150\times128\times256$ grid using 128 CPUs. Since we can run all ensemble members in parallel, the ensemble modeling we have performed here is feasible for real-time space weather forecasting. It is of importance to note that we have not considered the uncertainties in the input magnetic fluxes. Incorporating those uncertainties into the analysis may significantly increase our ensemble size. However, since the half-width and aspect ratio of the CME were shown to have minimal effect on the simulation results at Earth, we could still create ensembles of reasonable size by excluding those members.

\centerline{ACKNOWLEDGEMENT}
The authors acknowledge support from NASA/NSF SWQU grant 2028154. TKK acknowledges support from AFOSR grant FA9550-19-1-0027. NP was also supported, in part, by NSF-BSF grant 2010450 and NASA grants 80NSSC19K0075 and 80NSSC21K0004. Supercomputer allocations were provided on SGI Pleiades by NASA High-End Computing Program award SMD-21-44038581 and also on TACC Stampede2 and SDSC Expanse by NSF XSEDE project MCA07S033. We acknowledge the NASA/GSFC Space Physics Data Facility's OMNIWeb for the SW and IMF data used in this study. We also used SOHO and STEREO coronagraph data from https://lasco-www.nrl.navy.mil and stereo-ssc.nascom.nasa.gov respectively. SDO EUV and magnetogram data have been taken from http://jsoc.stanford.edu/ajax/exportdata.html. 
This work utilizes data produced collaboratively between the Air Force Research Laboratory (AFRL) and the National Solar Observatory.

\clearpage

\end{document}